\PassOptionsToPackage{numbers,sort&compress}{natbib}
\documentclass[final,3p,times]{elsarticle}

\usepackage{amssymb}
\usepackage{amsthm}
\usepackage{amsmath}
\usepackage{xfrac}
\usepackage{subcaption}

\usepackage[colorlinks]{hyperref}
\usepackage{cancel}
\usepackage{xcolor}
\usepackage{float} 
\usepackage[figurename=Fig.,labelfont=bf]{caption}

\usepackage[ruled,vlined]{algorithm2e} 
\makeatletter
\newcommand*{\algotitle}[2]{%
	\stepcounter{algocf}%
	\hypertarget{algocf.title.\theHalgocf}{}%
	\NR@gettitle{#1}%
	\label{#2}%
	\addtocounter{algocf}{-1}%
}
\makeatother
\usepackage{framed}
\usepackage{multicol}
\usepackage{cleveref}
\usepackage{hyperref}
\usepackage{longtable}
\usepackage[dvipsnames]{xcolor}
\usepackage{comment}
\usepackage{array}
\usepackage{multirow}
\usepackage{booktabs}
\usepackage{soul}
\usepackage{tabularx}

\immediate\write18{%
	makeindex -s nomencl.ist -o \jobname.nls -t \jobname.nlg \jobname.nlo%
}

\usepackage{nomencl}
\makenomenclature

\begin{document}
 
	\begin{frontmatter}
	
        \title{Machine learning enables roughness-driven inverse design of milling processes}
		
        \author[add1]{Hadi Bakhshan\fnref{fn1}}
        \author[add1,add2]{Sima Farshbaf\fnref{fn2}}
        \author[add1]{Fernando Rastellini}
        \author[add1,add3]{Josep Maria Carbonell}
        
        \fntext[fn1]{Email: hbakhshan@cimne.upc.edu}
        \fntext[fn2]{Email: sima.farshbaf@upc.edu}
        
		\address[add1]{Centre Internacional de Mètodes Numèrics a l'Enginyeria (CIMNE), Campus Norte UPC, 08034 Barcelona, Spain}
        \address[add2]{Universitat Politècnica de Catalunya (UPC), Campus Norte UPC, 08034 Barcelona, Spain}
 		\address[add3]{Mechatronics and Modelling Applied on Technology of Materials (MECAMAT) group. Universitat de Vic-Universitat Central de Catalunya (UVic-UCC), C. de la Laura 13, 08500 Vic, Spain}


\begin{abstract}

    Interest in applying data-driven approaches in manufacturing has grown significantly, particularly for mapping complex, high-dimensional relationships. The milling process is one area where predictive models can link influential parameters to surface roughness metrics prior to in situ operations. While this approach offers clear advantages, it faces challenges due to limited datasets and robustness issues in inverse design paradigms. To address these challenges, this paper proposes a machine learning (ML)-based framework for the inverse design of the surface milling process, with a focus on surface roughness as the design objective. The framework employs forward training of two ML models, a deep neural network (DNN) and a random forest (RF) ensemble, both developed using a high-fidelity synthetic dataset generated from a computational simulation framework. These trained models are integrated into a Bayesian optimization (BO) procedure to overcome the multiplicity problem arising from the many-to-one mapping inherent in the dataset. The approach identifies top-performing milling process configurations, considering both process and tool parameters, and presents them from the full solution space. The models achieve average relative errors below 5\% when compared to reference results, thereby demonstrating the robustness and reliability of the proposed methodology. 
    
\end{abstract}


		
	\end{frontmatter}
	


\section{Introduction}
	\label{introduction}

    The ability to transform raw materials into precise components by removing excess material makes machining a cornerstone of modern manufacturing. Among the various machining techniques, milling stands out for its flexibility, high productivity, and capability to achieve substantial material removal rates \cite{sun2023review}. It is widely applied across industries where both efficiency and precision are critical. 

    To ensure the precision and performance of machined components, surface quality is a critical factor, as the characteristics of a machined surface significantly influence functional properties such as friction \cite{berglund2010milled}, wear resistance \cite{bhushan2022effect, chang2024comprehensive}, contact behavior \cite{mu2019feasibility}, and fatigue life \cite{pegues2018surface}. Consequently, surface analysis metrics, particularly surface roughness, have become among the most widely used indicators in machining process design for evaluating surface integrity \cite{he2018influencing}. Accurate assessment of this parameter is essential to ensure the longevity and reliability of the final products.

    In surface-oriented machining process design, early prediction of surface roughness reduces trial-and-error efforts in identifying optimal machine and tool configurations. In modern manufacturing, machining process design increasingly begins with predefined surface roughness requirements, making it essential to establish reliable relationships between machining parameters and the resulting surface quality. To this end, predictive methods for surface roughness can be broadly classified into four categories. First, empirical models rely on experimental data and use regression techniques to relate cutting conditions, tool parameters, and material properties to roughness metrics \cite{he2018influencing}. However, they often suffer from limited generalizability and applicability, as extensive experimental campaigns are costly and time-consuming. Second, analytical methods describe surface formation through tool kinematics and mathematical formulations. They model the machined surface as the envelope of the cutter trajectory, enabling simplified estimation of roughness metrics \cite{sun2023review, yang2015surface}. However, these approaches are often insufficient for capturing complex machining conditions. Third, numerical approaches address these limitations by using discretization techniques to approximate analytical solutions across spatial and temporal domains. These methods are more flexible, allowing the incorporation of complex interactions between the tool and workpiece \cite{zhang2008new, xu20203d, wang2021modified}.

    Data-driven methods, such as machine learning (ML) and deep learning (DL), represent the fourth class of approaches. These methods rely on large datasets and are widely employed for surface roughness prediction \cite{yang2024review}. They incorporate tabular or image data obtained from surface analysis and use parameter-based models for feature extraction across different data types. Typically, these approaches learn a functional mapping between machining parameters, such as cutting speed, feed rate, cutting depth, and cutting width, and tool-related variables, such as tool diameter. Models including support vector machines (SVMs) \cite{abu2017surface, yeganefar2019use} and deep neural networks (DNNs) \cite{zain2010prediction, boga2021proper, wang2023knowledge}, among others, are commonly applied to tabular datasets.

    Beyond basic prediction, ML models contribute to deeper process understanding by identifying the relative importance and interactions among machining variables. This capability is particularly beneficial for process design and analysis. Such insights support process planning by enabling the forward design of machining conditions to meet target surface quality requirements. Furthermore, ML approaches can be enhanced through optimization techniques such as genetic algorithms (GA) \cite{mitchell1998introduction} and particle swarm optimization (PSO) \cite{kennedy1995particle}, which improve model structure, parameter tuning, and convergence behavior \cite{wang2022ensemble, li2021predicting, li2021effective}. These hybrid strategies can effectively mitigate common issues such as local minima and overfitting in standalone learning models.

    In addition, ML models can be extended by incorporating supplementary input signals, such as cutting forces, vibrations, acoustic emissions, and temperature. This leads to the development of multi-sensor fusion frameworks for real-time monitoring \cite{guo2023prediction, chen2024predicting, li2022application, li2022roughness}. Such integration enhances adaptability to dynamic, time-dependent data, which is inherent in machining processes, and provides a pathway toward intelligent manufacturing systems.

    Despite their strengths in modeling complex, high-dimensional relationships, ML methods have several limitations. Their performance is highly dependent on the quality and representativeness of the training dataset \cite{wang2022novel}, which often requires extensive and costly experimental campaigns or the generation of high-fidelity synthetic data. Additionally, challenges such as many-to-one mapping arise, particularly in the context of inverse design. 

    Inverse design, i.e., determining suitable machining or tool parameters given a target surface roughness, is of significant industrial importance, as it reduces costs and minimizes the need for extensive experimental trials. However, training ML models by simply swapping input and output variables can lead to averaged predictions when the dataset exhibits multiplicity or many-to-one mapping characteristics. In such cases, multiple combinations of input parameters correspond to similar or nearly identical output values. This phenomenon stems from complex and nonlinear interactions among machining variables, where compensatory effects exist. For instance, an increase in feed rate may degrade surface quality, but this effect can be offset by adjusting cutting speed or tool geometry, resulting in a similar roughness outcome.
    
    This many-to-one mapping introduces ambiguity in decision-making, particularly when the objective is to determine a unique or optimal set of machining conditions for a desired roughness target. While forward machine learning models for surface roughness prediction (inputs-to-output) are well established in the literature \cite{yang2024review, zeng2023milling, rifai2020evaluation, giusti2020image}, inverse design formulations (desired output-to-optimal inputs), particularly those addressing output multiplicity and non-uniqueness, remain comparatively underexplored.

    To address this limitation, coupling predictive ML models with optimization strategies is necessary. In such frameworks, ML models are used to learn the forward mapping, while optimization algorithms search the input space to identify parameter combinations that satisfy a target roughness value, potentially under additional constraints. Methods such as grid search \cite{liashchynskyi2019grid} and Bayesian optimization (BO) \cite{shahriari2015taking} are commonly used for solution space exploration \cite{selvarajan2024comprehensive}. Among these, BO offers a more efficient alternative by modeling the objective function probabilistically and iteratively selecting promising parameter sets.

    Therefore, in this study, we leverage ML and BO to perform inverse design and identify appropriate machining and tool parameters based on target surface roughness metrics. First, a large number of high-fidelity datasets are generated using forward solution methods (FSM) to simulate surface topography and extract average surface roughness and maximum height values (\cref{sec_computational_framework}). These simulations are conducted over a range of six input parameters, including cutting speed and feed rate (machining parameters), number of inserts, insert radius, and two installation error variables for an indexable face cutter. Subsequently, a multiplicity analysis is conducted to investigate the many-to-one mapping issue, thereby justifying the avoidance of direct inverse model training (\cref{sec_dataset_generation}). Based on recent benchmarking studies \cite{reddy2025performance}, two high-performing models including DNN and random forests (RF) are selected. These models are trained and optimized through hyperparameter tuning for both single-output and multi-output prediction scenarios (\cref{sec_machine_learning}). BO is then integrated to solve the inverse design problem. Finally, two case studies are presented, in which target surface roughness metrics are specified and the corresponding optimal parameter sets are identified (\cref{sec_inverse_design}). Performance evaluation against reference datasets demonstrates that the proposed inverse design framework robustly provides diverse design solutions with acceptable prediction errors below 5\%.


\section{Computational framework for synthetic dataset}
    \label{sec_computational_framework}

    Training ML methods requires large datasets to achieve accurate and reliable predictions. This need often leads to the use of synthetic data generation techniques. This section briefly discusses the computational framework used for dataset generation. A more extensive discussion, including detailed descriptions and validation procedures, can be found in \cite{bakhshan2026efficient}.

    \subsection{Framework}
	    \label{subsec_framework}

        Numerical methods are effective for modeling and predicting surface topography and roughness metrics in complex machining processes. In general, these methods rely on the discretization of spatial and temporal domains into finite intervals, enabling the progressive reconstruction of surface topography in a step-by-step manner.
        
        In this study, we employ forward solution method (FSM) to generate the dataset. FSM discretizes the cutting-edge, the workpiece, and time. At each time step, the motion of cutting-edge points within the workpiece material is tracked, and the resulting surface is obtained by assigning the minimum height value at each point on the workpiece grid. However, high computational costs are expected as the number of time steps and discretized points increases \cite{li2013surface, zhou2018surface}, necessitating the development of efficient approaches, particularly for generating large-scale datasets.
        
        The first step in FSM is to establish the kinematic equations of the cutting insert relative to the workpiece. This is achieved by defining appropriate coordinate systems for the cutting-edge and transforming them into the workpiece coordinate system to determine their spatial correspondence.
        
\begin{figure}[ht]%
            \centering
            \includegraphics[width=0.99\textwidth]{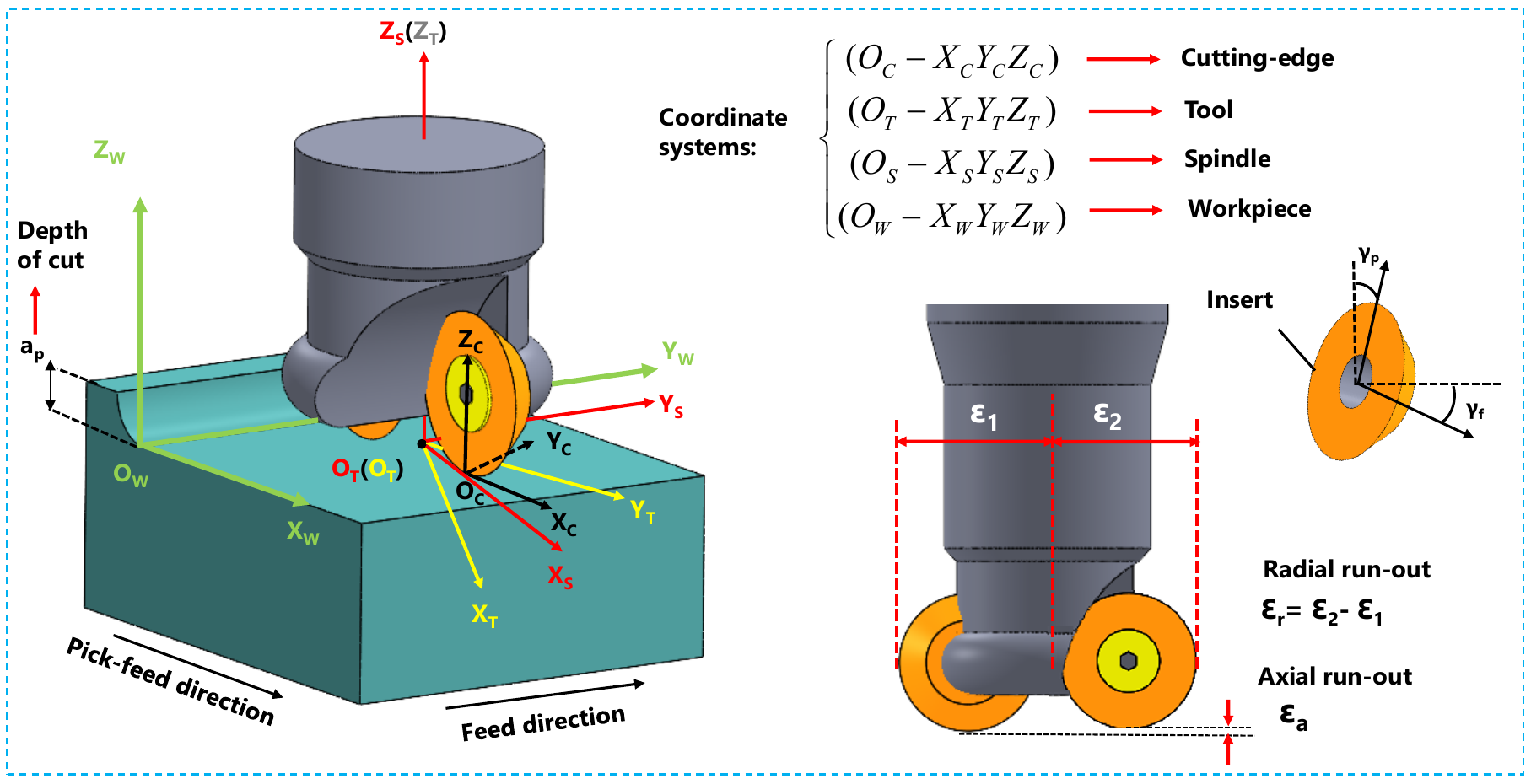}
            \caption{The face milling tool with two inserts interacting with the workpiece is illustrated, highlighting the four defined coordinate systems: cutting-edge, tool, spindle, and workpiece coordinate systems. The rake angles of the insert, along with the run-out parameters, are also presented.}\label{Tool_Insert_Coordinates}
        \end{figure}

        In this study, we focus on the face milling process with an indexable cutter, as schematically illustrated in \autoref{Tool_Insert_Coordinates}. For this problem, four coordinate systems can be defined: the cutting-edge coordinate system, the tool coordinate system, the spindle coordinate system, and the workpiece coordinate system, as shown in \autoref{Tool_Insert_Coordinates}. The final transformation equation from the cutting-edge coordinate system to the workpiece coordinate system for a given point on the cutting-edge can be expressed as

        \begin{equation}\label{eq_final_transformation}
            T_{C\to W} = \left[ \begin{matrix}
            {\cos(\gamma_f - \theta)} & {\cos(\gamma_f - \theta)\cos\gamma_p} & {\cos(\gamma_f - \theta)\sin\gamma_p} &
            {{{\varepsilon }_{r}}\cos(\gamma_f - \theta) + {{\varepsilon }_{a}}\sin(\gamma_f - \theta)\sin\gamma_p + x_0} \\
            {\sin(\gamma_f - \theta)} & {\sin(\gamma_f - \theta)\cos\gamma_p} & {\sin(\gamma_f - \theta)\sin\gamma_p} &
            {{{\varepsilon }_{r}}\sin(\gamma_f - \theta) + {{\varepsilon }_{a}}\cos(\gamma_f - \theta)\sin\gamma_p + y_0 + v_f t} \\
            0 & {-\sin\gamma_p} & {\cos\gamma_p} & {z_0 + {{\varepsilon }_{a}}\cos\gamma_p} \\
            0 & 0 & 0 & 1 \\
            \end{matrix} \right]
        \end{equation}

        \noindent where ${\gamma}_{f}$ is the radial rake angle and ${\gamma }_{p}$ denotes the axial rake angle of the tool. ${\varepsilon }_{r}$ and ${\varepsilon }_{a}$ represent the radial and axial run-outs of the (K)-th cutting-edge, respectively, as illustrated in \autoref{Tool_Insert_Coordinates}. The parameter ${v}_{f}$ denotes the feed speed, ${a}_{p}$ is the depth of cut, and $({{x}_{0}},{{y}_{0}},{{z}_{0}})$ represents the initial position of the tool in the workpiece coordinate system. The angular position $\theta$ is given by 

        \begin{equation}\label{eq_angular_position}
            \theta = \varphi + \frac{2\pi (K-1)}{z_n} - \omega t
        \end{equation}

        \noindent where $\varphi$ is the initial phase angle, $\omega$ is the angular velocity of the tool, ${z}_{n}$ denotes the number of tool teeth (inserts), and $t$ represents the milling time.

        Based on the kinematic equation given in \autoref{eq_final_transformation}, each discretized point on the cutting-edge is evaluated at every time step to determine its position in the workpiece coordinate system. This procedure requires nested loop operations in the algorithm, which introduce significant computational overhead when large numbers of spatial and temporal elements are considered. However, these limitations can be effectively mitigated through parallelization and compilation techniques, as discussed in \cite{bakhshan2026efficient}. The implementation of computational framework is available on GitHub at \url{https://github.com/HadiBakhshan/surf-topo.git}, along with comprehensive documentation and detailed compilation guidelines.

    \subsection{Validation}
	    \label{subsec_validation}

        To validate the computational model for dataset generation, experimental results from a face milling process conducted on HT300 grey cast iron workpieces, as reported in \cite{wang2023high}, are utilized. \autoref{tab_tool_paramters} summarizes the tool parameters used in the experiments. Additionally, \autoref{tab_milling_paramters} presents three case studies, including the corresponding milling parameters and their associated run-outs evaluated from the measurements \cite{wang2023high}.

        \begin{table}[!t]
            \centering
            \fontsize{8}{13}\selectfont
            \caption{Tool parameters \cite{wang2023high}.}
            \label{tab_tool_paramters}
            \begin{tabularx}{0.5\textwidth}{X X}
                \toprule
                Parameter & Value \\ \toprule
    
                Tool & R217.29-2520.3-05.2.070 \\
    
                Cutting diameter & 10.0 mm \\
    
                Cutting diameter maximum & 20.0 mm \\
    
                Shank length & 66.0 mm \\
    
                Shank diameter & 25.0 mm \\
    
                Radial rake angle & ${{0.6}^{\circ }}$ \\
    
                Axial rake angle & ${{0.0}^{\circ }}$ \\
    
                Tooth number & 2 \\
    
                Insert & RDHW10T3M0-8-MD04 F40M \\
    
                Insert thickness & 3.97 mm \\
    
                Insert diameter & 10.0 mm \\
    
                Corner radius & 5.0 mm \\
    
                Coating technology & PVD \\
       
                \bottomrule
            \end{tabularx}
        \end{table}

        \begin{table}[!t]
            \centering
            \fontsize{8}{13}\selectfont
            \caption{Milling process parameters with measured run-outs \cite{wang2023high}.}
            \label{tab_milling_paramters}
            \begin{tabularx}{0.8\textwidth}{>{\raggedright\arraybackslash}X 
                                    >{\raggedright\arraybackslash}X 
                                    >{\raggedright\arraybackslash}X
                                    >{\raggedright\arraybackslash}X
                                    >{\raggedright\arraybackslash}X 
                                    >{\raggedright\arraybackslash}X}
                \toprule
                No. & ${{v}_{c}}$ (m/min) & ${{f}_{z}}$ (mm/tooth) & ${{a}_{P}}$ (mm) & ${{\varepsilon }_{r}}$ (mm) & ${{\varepsilon }_{a}}$ (mm)  \\ \toprule
    
                Case 1 & 170 & 0.6 & 0.5 & +0.011 & +0.003 \\
    
                Case 2 & 200 & 0.4 & 0.4 & $-0.026$ & +0.009 \\
    
                Case 3 & 230 & 0.5 & 0.3 & $-0.013$ & +0.007 \\
       
                \bottomrule
            \end{tabularx}
        \end{table}

\begin{figure}[!t]%
            \centering
            \includegraphics[width=0.99\textwidth]{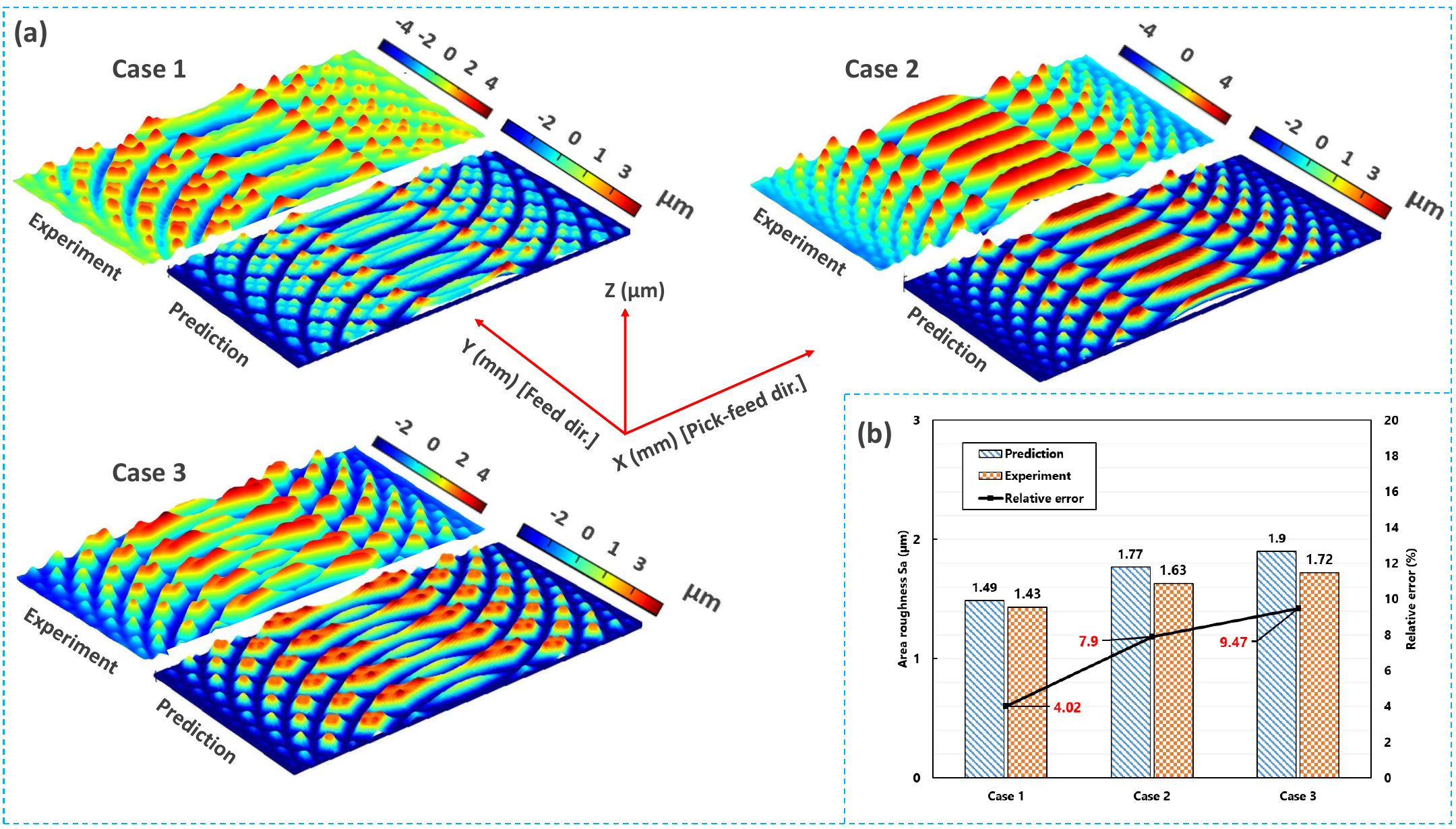}
            \caption{(a) 3D surface topographies from both model predictions and experiments for the three case studies. (b) Comparison between predicted and experimental average surface roughness values, along with the corresponding relative errors. Experimental data are adapted from \cite{wang2023high}.}\label{3D_Surface_Topography}
        \end{figure}

        \autoref{3D_Surface_Topography}a depicts the surface topography generated by the computational model for the case studies considered. It is evident that the model successfully captures the main characteristics of the machined surfaces, showing strong agreement with the experimental results in terms of both surface morphology and contour distribution.
        
        The formation of surface irregularities, such as peaks and valleys, is primarily governed by the alternating engagement of the tool during forward and backward cutting actions, considering the presence of two inserts in this specific scenario. These characteristic features can be directly associated with the cutting-edge trajectories. Each cutting-edge follows a distinct path, and the combined effect of tool rotation and feed motion determines the resulting surface topography. Regions with sparse or absent cutting-edge trajectories tend to exhibit more pronounced peaks, whereas areas with a higher density of trajectory intersections generally correspond to lower surface heights.
        
        In addition to the qualitative assessment provided by the surface topography, quantitative parameters offer deeper insight into the predictive capacity of the model and its suitability to meet surface quality design requirements. Among these parameters, the most widely used metric is the arithmetic mean height, or average surface roughness ($S_a$), which represents the mean of the absolute deviations of the surface heights from the reference plane and is defined as

        \begin{equation}\label{eq_average_roughness}
            {{S}_{a}}=\frac{1}{A}\iint\limits_{A}{|z(x,y)|dA}
        \end{equation}

        \noindent where $A$ is the measured surface area and $z(x,y)$ denotes the height deviation from the mean plane. The maximum surface height, ($S_z$) defined as the vertical distance between the deepest valley and the highest peak, is another practical and widely used parameter. Together with $S_a$, it provides a more comprehensive characterization of surface quality and is given by

        \begin{equation}\label{eq_max_roughness}
            {{S}_{z}}={{z}_{\max }}-{{z}_{\min }}
        \end{equation}

        These two parameters are particularly meaningful in the context of machining design criteria, as they align well with the practical intuition of designers. In most cases, designers specify surface quality requirements in terms of allowable roughness limits. Therefore, by defining target values for average roughness or maximum permissible surface height, the design process can be approached inversely enabling the determination of the necessary process parameters to achieve the desired surface quality. 

        With regard to average surface roughness, \autoref{3D_Surface_Topography}b presents a comparison between predicted and experimental values across the three case studies. The results show encouraging agreement across the three validation cases considered, with an average relative error of approximately 7\%, indicating that the framework successfully captures the dominant surface roughness trends under the tested conditions. The reliability of the computational framework is further demonstrated through additional experimental results reported in \cite{bakhshan2026efficient}.

\section{Dataset generation}
    \label{sec_dataset_generation}

    \subsection{Inputs and outputs}
	    \label{subsec_inputs_outputs}

\begin{figure}[!t]%
            \centering
            \includegraphics[width=0.7\textwidth]{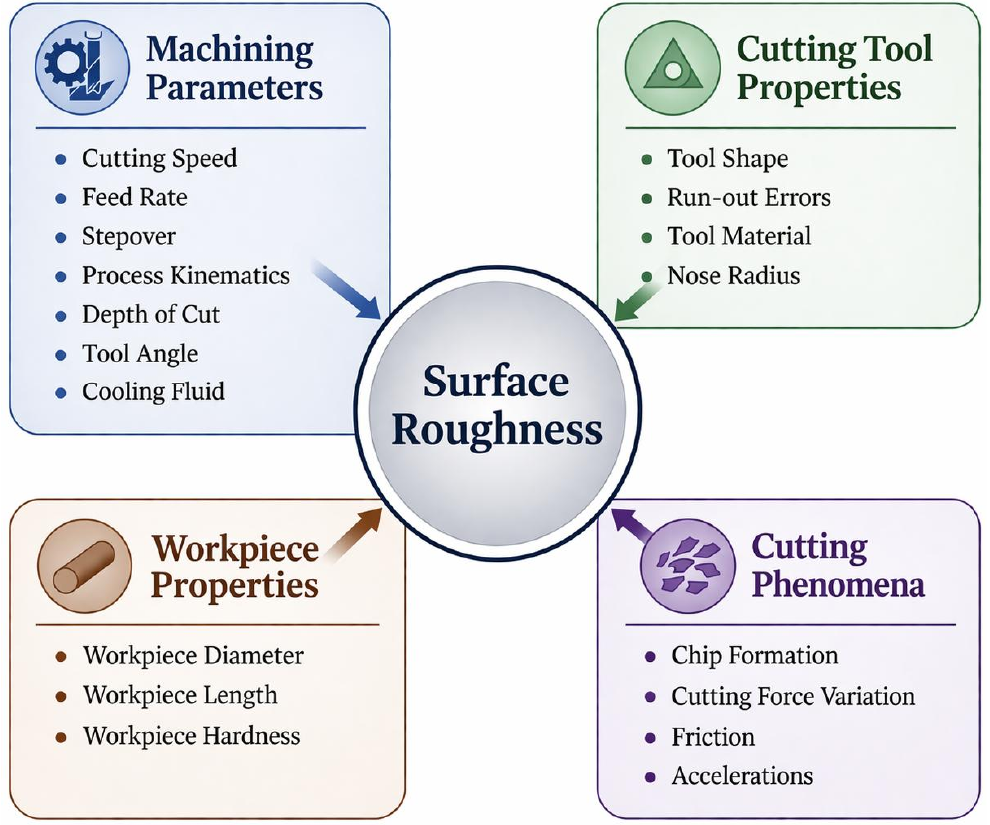}
            \caption{Parameters that influence the surface roughness in machining \cite{benardos2003predicting}.}\label{Roughness_Influence_Parameters}
        \end{figure}

        Surface roughness can be considered the target metric for inverse design in milling and is influenced by multiple factors, which can be classified into four main categories: machining conditions, tool specifications, workpiece material properties, and the resulting cutting phenomena, as illustrated in \autoref{Roughness_Influence_Parameters}.

        Given the computational model for dataset generation, which is a geometrical model that disregards material properties and treats all components as rigid bodies, it is evident from \autoref{eq_final_transformation} that the influential parameters in the calculations are limited to milling parameters and tool specifications.
    
        Among the milling parameters, the primary factors affecting surface roughness are the cutting speed (${v}_{c}$) and feed rate (${f}_{z}$). From the tool perspective, the parameters investigated for their influence on surface roughness include tool geometry (shape) and run-out errors (${{\varepsilon }_{r}}$ and ${{\varepsilon }_{a}}$). In this study, for the specific indexable tool highlighted in \autoref{tab_tool_paramters}, the tool geometry is defined by the number of inserts ($z$) and their radius (${{r}_{i}}$).
        
        For dataset generation, all relevant parameters influencing the milling process and tool characteristics are summarized in \autoref{tab_range_paramters}, along with their corresponding ranges. These ranges are selected based on the process requirements and design specifications.
        
        Continuous parameters, such as cutting speed (${v}_{c}$), feed rate (${f}_{z}$), radial run-out (${\varepsilon}_{r}$), and axial run-out (${\varepsilon}_{a}$), are defined by minimum and maximum values from which samples are drawn. The ranges of ${v}_{c}$ and ${f}_{z}$ are selected to cover a broad spectrum of milling conditions, with reference to the case studies highlighted in \autoref{tab_milling_paramters}. Additionally, the ranges for ${\varepsilon}_{r}$ and ${\varepsilon}_{a}$ are determined based on experimentally measured minimum and maximum values reported in \autoref{tab_milling_paramters}.
        
        Discrete parameters include the number of inserts and the insert radius. For the specific tool type considered in this study, tools with 2, 3, and 4 inserts, along with their corresponding radii, are included, as specified in the manufacturing catalog.

        \begin{table}[!t]
            \centering
            \fontsize{8}{13}\selectfont
            \caption{Range of parameters for dataset generation.}
            \label{tab_range_paramters}
            \begin{tabularx}{0.8\textwidth}{
                >{\raggedright\arraybackslash}X 
                >{\raggedright\arraybackslash}X 
                >{\raggedright\arraybackslash}X 
                >{\raggedright\arraybackslash}X}
                        
                \toprule
                 & Parameter  & Type & Range \\ \toprule
    
                \multirow[t]{2}{=}{Milling process} & Cutting speed ${{v}_{c}}$ (m/min) & Continuous & 100 -- 300 \\
    
                & Feed rate ${{f}_{z}}$ (mm/tooth) & Continuous & 0.1 -- 0.9 \\
    
                \cmidrule{1-4}
    
                \multirow[t]{2}{=}{Tool} & Number of inserts $z$ & Discrete &  2, 3, 4 \\
    
                & Insert radius ${{r}_{i}}$ (mm) & Discrete &  3, 4, 5 \\
    
                & Radial run-out ${{\varepsilon }_{r}}$ (mm) & Continuous &  -0.026 -- 0.011 \\
    
                & Axial run-out ${{\varepsilon }_{a}}$ (mm) & Continuous &  0.003 -- 0.009 \\
    
                \bottomrule
            \end{tabularx}
        \end{table}

        To generate a dataset consisting of 10,000 samples from the space of possible parameter combinations, Latin Hypercube Sampling (LHS) was employed as a random sampling strategy, using a seed value of 42. The sampling was performed within the ranges of the input parameters specified in \autoref{tab_range_paramters}, while all other model parameters were kept constant, as defined previously. 
    
        The output (target) parameters considered in this study are the average surface roughness, $S_a$, and the maximum surface height, $S_z$. For each simulation run, the model computes the surface topography over a workpiece region of interest measuring $10\times 5\,m{{m}^{2}}$, taking into account the engagement of the tool inserts after a stable feed motion of the workpiece. Subsequently, the values of $S_a$ and $S_z$ are calculated from the generated surface and stored as output data.

        \begin{figure}[!t]%
            \centering
            \includegraphics[width=0.88\textwidth]{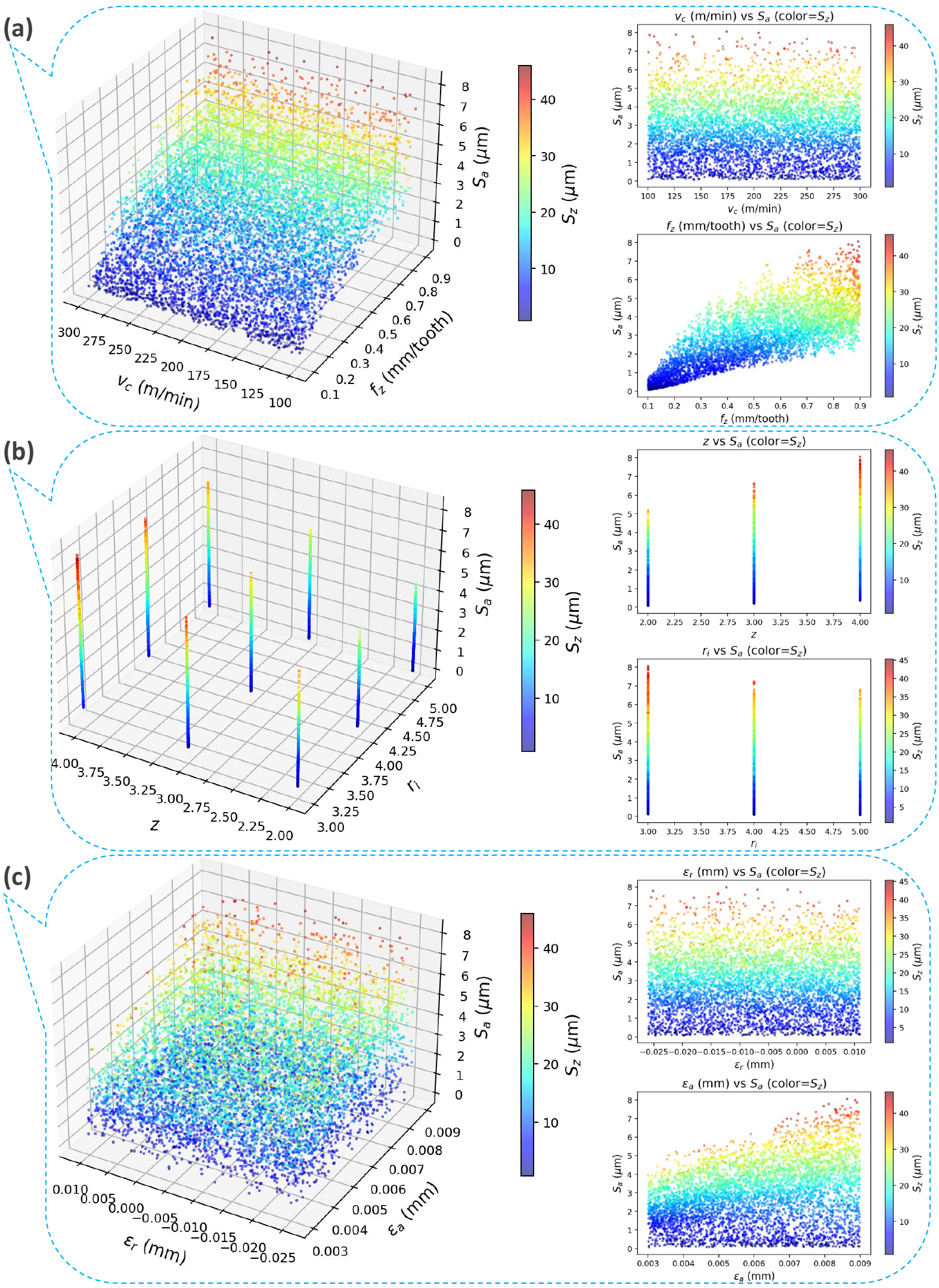}
            \caption{3D and 2D diagrams showing the variation of outputs such as average surface roughness and maximum surface height with respect to six input parameters, namely: (a) cutting speed and feed rate, (b) number of inserts and their radii, and (c) radial and axial run-outs.}\label{Dataset_Variation}
        \end{figure}

    \subsection{Dataset analysis}
	    \label{subsec_dataset_analysis}
        
        This section examines the effect of variations in the input parameters on the outputs, namely $S_a$ and $S_z$. To achieve this, the generated dataset is utilized such that one parameter is varied while the remaining parameters are kept constant. \autoref{Dataset_Variation} illustrates the variation of the six input parameters with respect to the two outputs.

        For cutting speed, the dataset appears to be approximately uniformly distributed across the design space as shown in \autoref{Dataset_Variation}a. Both low and high values of cutting speed correspond to a wide range of average and maximum roughness values. This indicates that lower values of $S_a$ and $S_z$ can occur at both low and high cutting speeds. Similarly, higher roughness values are also observed across the entire range of cutting speeds, suggesting that there is no strong direct dependency between cutting speed and the outputs.

        In contrast, the behavior observed for feed rate is markedly different. The corresponding plot (see \autoref{Dataset_Variation}a) demonstrates a clear direct relationship: as the feed rate increases, both average and maximum roughness values increase. The data points are concentrated toward higher output values, and the color variation further emphasizes the increase in maximum roughness. Therefore, higher feed rates are associated with higher surface roughness.
        
        \autoref{Dataset_Variation}b shows the variation of outputs with respect to discrete input parameters, including the number of inserts and insert radius. It is evident that increasing both parameters affects the outputs; however, their influence differs in magnitude and trend. For the number of inserts, tools with two inserts exhibit lower maximum values of both average and maximum roughness compared to tools with four inserts. Conversely, the trend is reversed for insert radius: increasing the radius from 3 mm to 5 mm results in lower values of both outputs.

        Finally, the dependency of the outputs on run-out parameters is illustrated in \autoref{Dataset_Variation}c. For radial run-out, the dataset shows a distribution and color pattern similar to that of cutting speed, indicating a weak or non-distinct relationship with the outputs. However, axial run-out exhibits a clearer trend: lower values of axial run-out correspond to lower roughness values, while increasing axial run-out leads to higher values of both $S_a$ and $S_z$.

        In general, the dependency of the outputs on each input parameter is analyzed in this section. Nonlinear relationships are particularly evident for parameters such as feed rate and axial run-out. These dependencies become increasingly complex as the input space expands from one-dimensional to six-dimensional. Consequently, mapping the inputs to the outputs becomes more challenging, especially in inverse design problems, where the goal is to determine appropriate input parameters that yield desired output values, such as specific roughness metrics.

    \subsection{Multiplicity analysis}
	    \label{subsec_multiplicity_analysis}

        To further investigate the relationship between the process parameters and the resulting surface characteristics, an exploratory analysis was conducted on the dataset. The primary objective of this analysis is to assess the uniqueness of the input–output mapping, with particular emphasis on identifying potential many-to-one relationships, where distinct combinations of input parameters yield similar outputs.

        To this end, clustering was performed in the output space using the density-based spatial clustering of applications with noise (DBSCAN) algorithm \cite{schubert2017dbscan}. DBSCAN groups data points based on spatial density, enabling the identification of regions of similar outputs without requiring prior specification of the number of clusters. A distance-based tolerance parameter defines the maximum separation between points within the same cluster, while points in low-density regions are treated as noise.

        Clustering was applied using a tolerance of 0.01 to both individual output dimensions and joint output space. The resulting cluster assignments allow for the identification of groups of samples that produce nearly identical surface characteristics. The following results are presented for the joint output space.

\begin{figure}[!t]%
            \centering
            \includegraphics[width=0.99\textwidth]{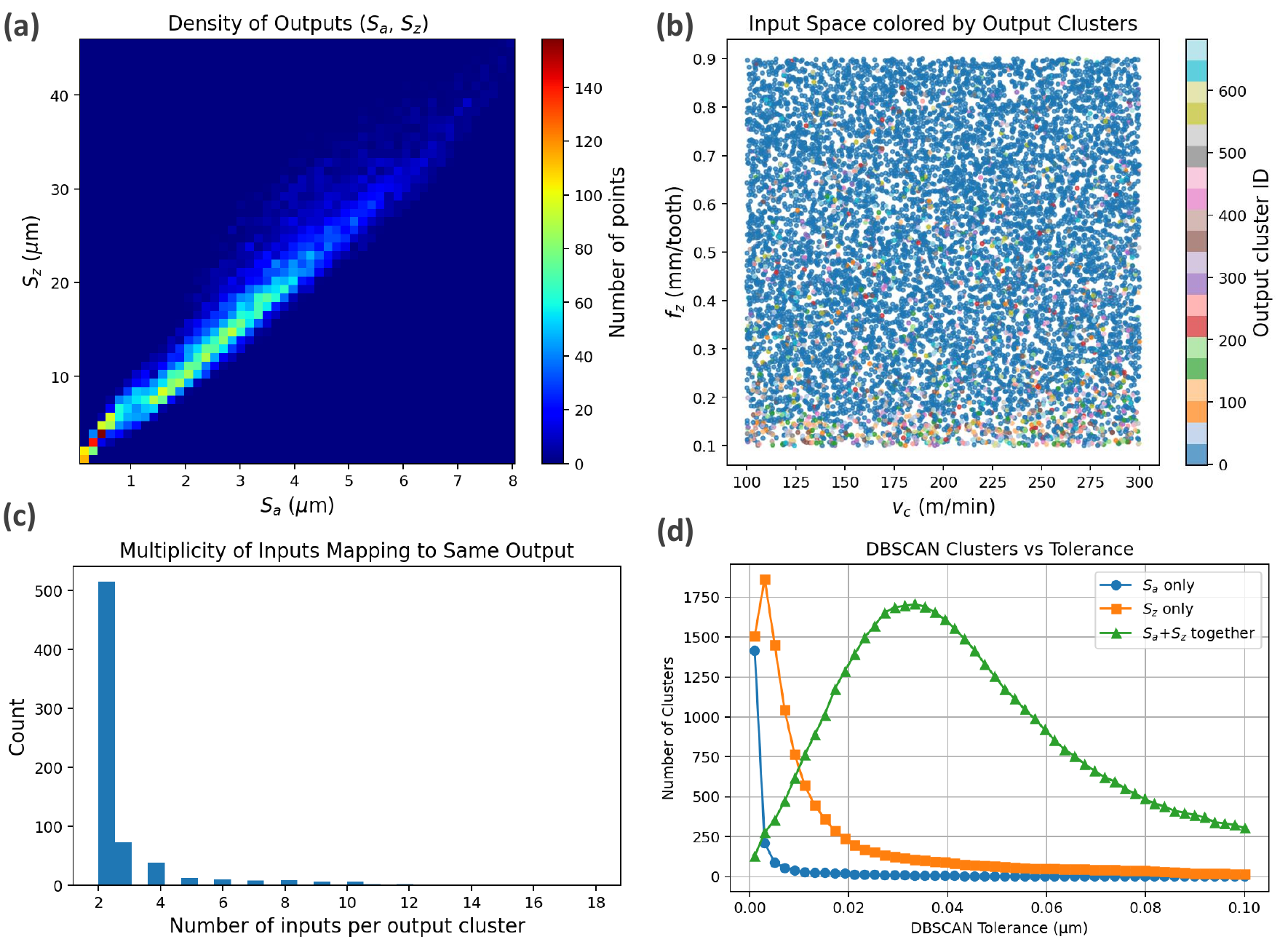}
            \caption{(a) Density distribution of the output space. (b) Projection of output clusters onto the input space defined by cutting speed and feed rate. (c) Multiplicity distribution as a function of the number of input variables involved. (d) Sensitivity of clustering results to tolerance parameters for each output configuration.}\label{Dataset_Multiplicity}
        \end{figure}

        \autoref{Dataset_Multiplicity}a shows a two-dimensional histogram of average roughness versus maximum surface height, used to visualize the distribution and concentration of surface roughness metrics. It can be observed that a large portion of the dataset is concentrated at relatively low values of both outputs.

        \autoref{Dataset_Multiplicity}b illustrates the projection of output clusters onto the input space. In this case, the input parameters of cutting speed and feed rate are visualized, with data points colored according to their assigned output cluster. This mapping provides insight into whether similar outputs originate from localized or dispersed regions in the input space. Notably, certain output clusters are associated with input samples that are widely distributed across the parameter space. This indicates that similar surface characteristics can be obtained from substantially different process conditions, highlighting a pronounced many-to-one mapping and non-uniqueness in the input–output relationship.

        For each output cluster, \autoref{Dataset_Multiplicity}c shows the number of associated input samples. This metric quantifies the degree of non-uniqueness in the mapping: large clusters indicate that multiple distinct input configurations produce similar outputs, while small clusters suggest more deterministic behavior. In this case, many clusters correspond to only two input samples, indicating limited redundancy in most regions of the output space.

        To evaluate the sensitivity of the clustering results to the tolerance parameter, the number of clusters obtained using DBSCAN was analyzed as a function of the distance threshold, varied from 0.001 to 0.1 $\mu$m. \autoref{Dataset_Multiplicity}d shows that, for average roughness, the number of clusters decreases rapidly with increasing tolerance and quickly converges to very low values, indicating low variability and strong concentration. In contrast, maximum surface height exhibits an initial increase in the number of clusters at small tolerances, followed by a gradual decrease and convergence, reflecting a more complex and dispersed distribution. For the joint output space, the number of clusters initially increases, then decreases, and eventually stabilizes around approximately 300 clusters. This behavior reflects an initial resolution of fine-scale structure followed by cluster merging at larger tolerances.

        Overall, the results indicate that average roughness varies weakly, whereas maximum surface height plays a dominant role in structuring the output space. This leads to persistent many-to-one relationships between inputs and outputs, highlighting the inherent non-uniqueness of the process mapping. The implementation of the dataset analysis was carried out using Python with the \texttt{scikit-learn} library and is available on GitHub at \url{https://github.com/HadiBakhshan/roughness-inverse-ml.git}.

\section{Machine learning (ML) model}
    \label{sec_machine_learning}

    ML models are typically employed to learn a mapping from input parameters to output responses, a formulation commonly referred to as the forward design problem. However, by swapping the input and output parameters, the problem can be reformulated as an inverse design problem. This capability is generally lacking in conventional computational and numerical modeling approaches, where machining parameters (inputs) are used solely to predict surface roughness (outputs).
        
    However, in the case of surface roughness prediction, this approach becomes problematic. The output space (e.g., roughness metrics) is relatively limited in range, while the number of influential input parameters (such as machining and tool parameters) is large. As a result, multiple distinct combinations of input parameters can produce similar or nearly identical output values. This leads to a many-to-one relationship in the forward mapping.
    
    When the mapping is inverted by simply exchanging inputs and outputs, the problem becomes one-to-many, which is inherently ill-posed. Training a standard regression model under these conditions forces the model to approximate multiple valid input configurations with a single averaged prediction. Consequently, the model tends to produce mean-valued outputs that may not correspond to any physically meaningful or feasible set of machining parameters.
    
    This averaging effect leads to erroneous or suboptimal recommendations for design parameters, thereby limiting the reliability of such inverse models. Consequently, inverse design in this context typically requires more advanced methodologies such as probabilistic modeling, multi-modal prediction, or constraint-based optimization to properly address the inherent non-uniqueness of the input-output relationship. However, given the nature of the present problem, which demands physical consistency, deterministic behavior, and involves a lack of diverse data modalities, probabilistic and multi-modal approaches may be inadequate or ill-suited. Therefore, optimization-based approaches that incorporate forward ML models are preferred.

    Among ML models, deep learning (DL) approaches are widely recognized as highly effective for surface roughness prediction \cite{yang2024review}. In particular, deep neural networks (DNNs) provide a supervised learning framework well-suited for numerical data. In addition to DNNs, random forests (RF), as an ensemble learning method, have been reported to outperform other approaches in predicting surface roughness \cite{reddy2025performance}. Therefore, in this study, we develop both DNN and RF models for an inverse design framework, while noting that other regression-based models can also be incorporated as needed. The implementation of the models was carried out using \texttt{pytorch} and is available on GitHub at \url{https://github.com/HadiBakhshan/roughness-inverse-ml.git}.

    \subsection{Deep neural network (DNN) model}
	    \label{subsec_deep_neural_network_model}

        \subsubsection{DNN architecture}
            \label{subsubsec_DNN_architecture}

            To approximate the mapping between a 6-dimensional input space (${{v}_{c}}$, ${{f}_{z}}$, $z$, ${{r}_{i}}$, ${{\varepsilon }_{r}}$, ${{\varepsilon }_{a}}$) and a 2-dimensional output space ($S_a$, $S_z$) in one scenario, and a 1-dimensional output ($S_a$ only) in another, a fully connected DNN was implemented within a forward modeling framework. The dataset, consisting of 10,000 samples, was randomly split into training (80\%) and validation (20\%) subsets. Both input and output variables were standardized to have zero mean and unit variance, ensuring numerical stability and improving the conditioning of the optimization problem.

\begin{figure}[!t]%
                \centering
                \includegraphics[width=0.99\textwidth]{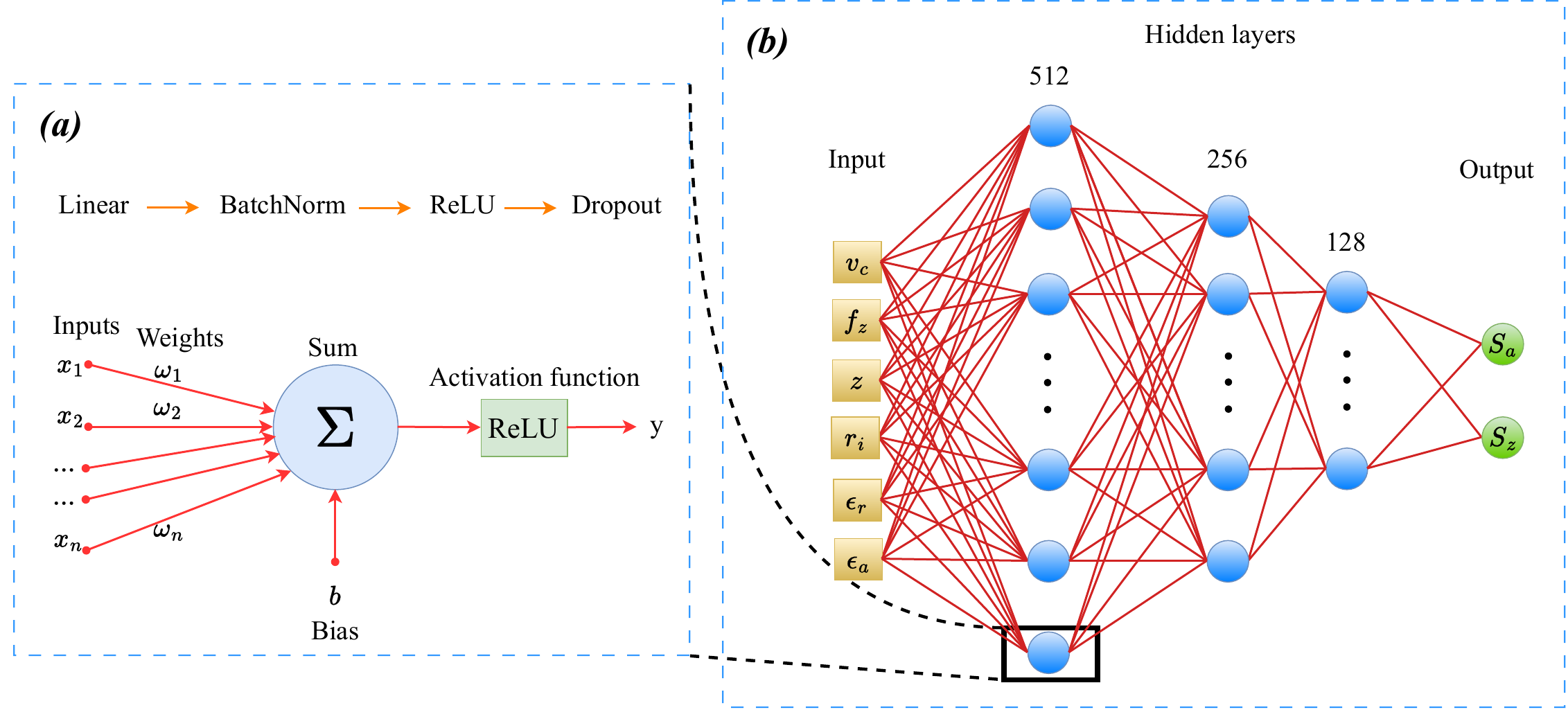}
                \caption{(a) Schematic representation of a node, including inputs, weights, bias, and the ReLU activation function. (b) Schematic of a DNN with milling process and tool parameters as inputs, three hidden layers with progressively decreasing width, and average roughness and maximum height as outputs.}\label{DNN_Schematic}
            \end{figure}

            The DNN architecture follows a funnel-shaped design, in which the number of neurons decreases with depth. This structure enforces a progressive compression of feature representations, enabling the network to learn compact latent representations of the input-output relationship. As illustrated in \autoref{DNN_Schematic}a, each hidden layer applies a linear transformation followed by batch normalization, which reduces internal covariate shift and stabilizes training, a ReLU activation function, and dropout. Dropout introduces stochastic regularization by randomly deactivating nodes during training, thereby reducing co-adaptation and improving generalization.

            The formulated model is cast as a nonlinear regression problem, with the objective of minimizing the mean squared error (MSE) between predicted and target outputs. Optimization is performed using the AdamW algorithm, which combines adaptive moment estimation with decoupled weight decay \cite{loshchilov2017decoupled}. The inclusion of weight decay acts as an $L_2$ regularization term, penalizing large weights and thereby controlling model complexity.

            To mitigate overfitting, early stopping is employed based on validation loss, with a patience of 20 epochs. This ensures that training terminates once performance on unseen data ceases to improve, effectively selecting a model near the optimal point of the bias-variance trade-off.

        \subsubsection{DNN hyperparameter optimization}
            \label{subsubsec_DNN_hyperparameter}

            To identify the most suitable model configuration, hyperparameters were optimized using Optuna \cite{akiba2019optuna}, a gradient-free framework for automated hyperparameter search, instead of relying on manual tuning. \autoref{tab_hyperparameters} lists the search space for key hyperparameters, including network width (number of nodes), depth (number of hidden layers), dropout rate, learning rate, weight decay, and batch size. These parameters directly influence model capacity, training dynamics, and regularization strength.

            \begin{table}[h]
                \centering
                \fontsize{8}{13}\selectfont
                \caption{Range of hyperparameters explored for determining the optimal DNN architecture.}
                \label{tab_hyperparameters}
                \begin{tabularx}{0.8\textwidth}{
                    >{\raggedright\arraybackslash}X
                    >{\raggedright\arraybackslash}X
                    >{\raggedright\arraybackslash}X
                    >{\raggedright\arraybackslash}X}
                
                    \toprule
                    Hyperparameter & Range / Values & Optimized value (two-output) & Optimized value (single-output) \\ \toprule
        
                    Width & 64, 128, 256, 512 & 512 & 512 \\
        
                    Depth & 2, 3, 4, 5, 6, 7 & 3 & 3 \\
        
                    Dropout & 0 -- 0.4 & 0.026 & 0.17 \\
        
                    Learning rate (lr) & Log[$1\times {{10}^{-4}}$, $3\times {{10}^{-3}}$] & $7.79\times {{10}^{-4}}$ & $7.79\times {{10}^{-4}}$ \\
        
                    Weight decay & Log[$1\times {{10}^{-6}}$, $1\times {{10}^{-3}}$] & $2.09\times {{10}^{-5}}$ & $1.53\times {{10}^{-5}}$ \\
        
                    Batch size & 32, 64, 128, 256 & 128 & 64 \\
         
                    \bottomrule
                \end{tabularx}
            \end{table}

\begin{figure}[!t]%
                \centering
                \includegraphics[width=0.99\textwidth]{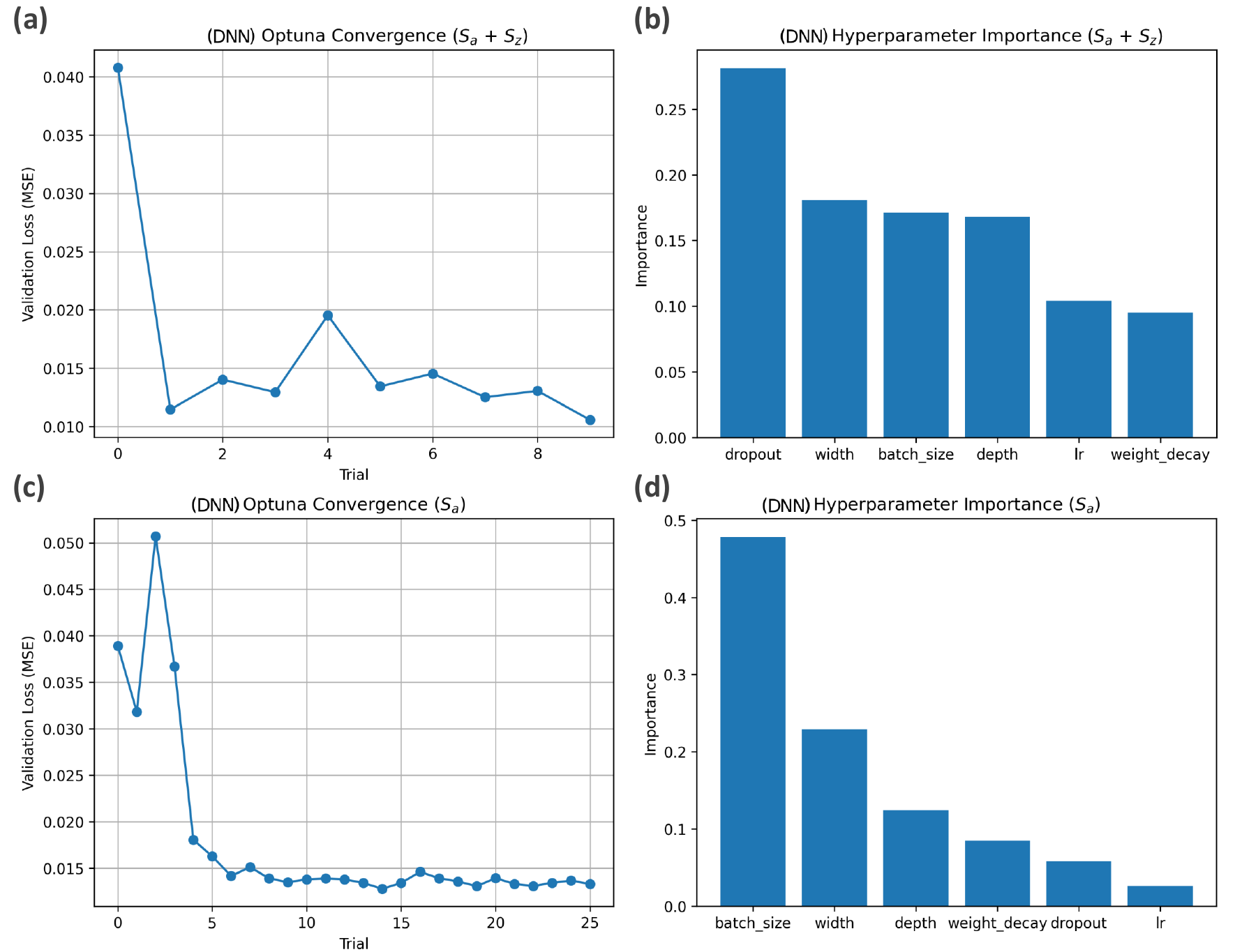}
                \caption{(DNN) Convergence of the hyperparameter optimization process over successive trials for (a) two-output scenario and (c) one-output scenario. Relative contribution of each hyperparameter to the observed variation in validation loss for (b) two-output scenario and (d) one-output scenario.}\label{Hyperparameter_Importance_DNN}
            \end{figure}

            The optimization proceeds sequentially to minimize the validation MSE: for each trial, a set of hyperparameters is sampled, a model is trained, and its validation loss is recorded. A median pruning strategy is employed to terminate underperforming trials early based on intermediate validation results. This pruning strategy is clearly reflected in the results: for the two-output scenario, approximately 80\% of trials are pruned (see \autoref{Hyperparameter_Importance_DNN}a), while for the single-output scenario, nearly 50\% are pruned (see \autoref{Hyperparameter_Importance_DNN}c) after only a few epochs, leading to a significant reduction in computational cost.

            The final best-performing hyperparameters for both scenarios are listed in \autoref{tab_hyperparameters}, and the corresponding network architecture, which has an identical number of layers and nodes, is illustrated in \autoref{DNN_Schematic}b. The DNN comprises three hidden layers with 512 nodes in the first layer, 256 nodes in the second, and 128 nodes in the third. For the two-output scenario, this configuration achieved a minimum validation MSE of 0.01057, as shown in \autoref{Hyperparameter_Importance_DNN}a, while for the single-output scenario, the minimum validation MSE was 0.0127 as shown in \autoref{Hyperparameter_Importance_DNN}c. These results indicate that shallow networks with three layers consistently outperformed deeper networks of up to six layers, suggesting that the dataset complexity is adequately captured with moderate depth and sufficiently wide layers.

            Furthermore, \autoref{Hyperparameter_Importance_DNN}b and \autoref{Hyperparameter_Importance_DNN}d quantify the contribution of each hyperparameter to the observed variation in validation loss using a variance-based decomposition method for the two-output and single-output scenarios, respectively. For the two-output scenario, dropout emerges as the dominant factor, while other parameters, including layer width, batch size, and depth, exhibit comparable influence on network performance. For the single-output scenario, batch size is the predominant factor, with the optimal value of 64 obtained through the optimization.

        \subsubsection{DNN model performance}
            \label{subsubsec_DNN_performance}

            The best-performing configuration for both scenarios was selected to train the model using the corresponding hyperparameters highlighted in \autoref{tab_hyperparameters}. The model was trained over multiple epochs, performing forward passes to generate predictions, computing the weighted loss, backpropagating gradients, and updating parameters via the AdamW optimizer. The performance of the predictions was evaluated after each epoch on the validation set to monitor generalization. Early stopping with checkpoints was employed: if the validation loss did not improve for a predefined number of epochs, training was terminated, and the best-performing model was saved.

\begin{figure}[!t]%
                \centering
                \includegraphics[width=0.99\textwidth]{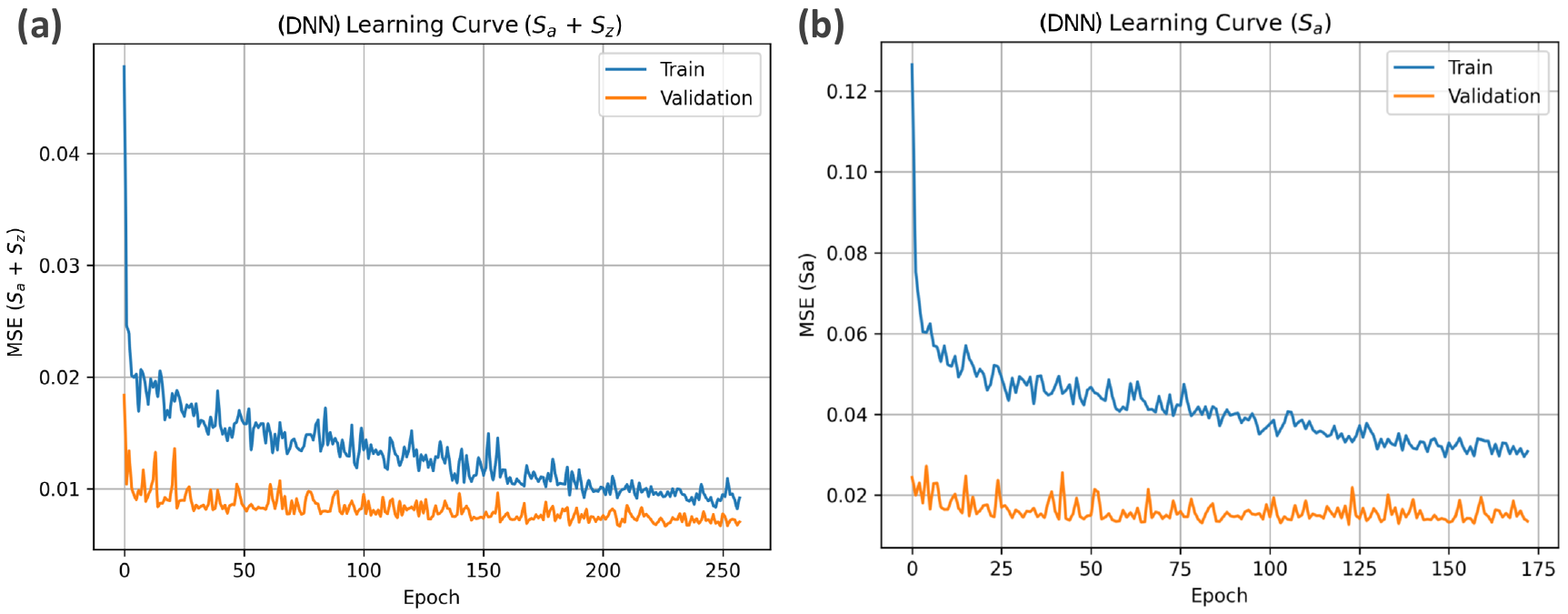}
                \caption{(DNN) Learning curves showing the training and validation loss for (a) the two-output scenario and (b) the single-output scenario.}\label{Learning_Curves_DNN}
            \end{figure}

            \autoref{Learning_Curves_DNN} illustrates the learning curves for both scenarios. The learning curves demonstrate the convergence of the training and validation losses toward stable MSE values. For the two-output scenario, as shown in \autoref{Learning_Curves_DNN}a, both the training and validation curves converge to MSE values below 0.01. In the single-output scenario, as shown in \autoref{Learning_Curves_DNN}b, the validation set achieves an MSE below 0.02, while the training set converges to approximately 0.03. The smooth decrease of both training and validation losses over epochs indicates stable convergence without overfitting, which is further supported by the early stopping mechanism. These low errors indicate that the DNN model accurately captures the relationship between the input features and the target surface roughness parameters in both scenarios, although the single-output scenario exhibits slightly higher error.

\begin{figure}[!t]%
                \centering
                \includegraphics[width=0.9\textwidth]{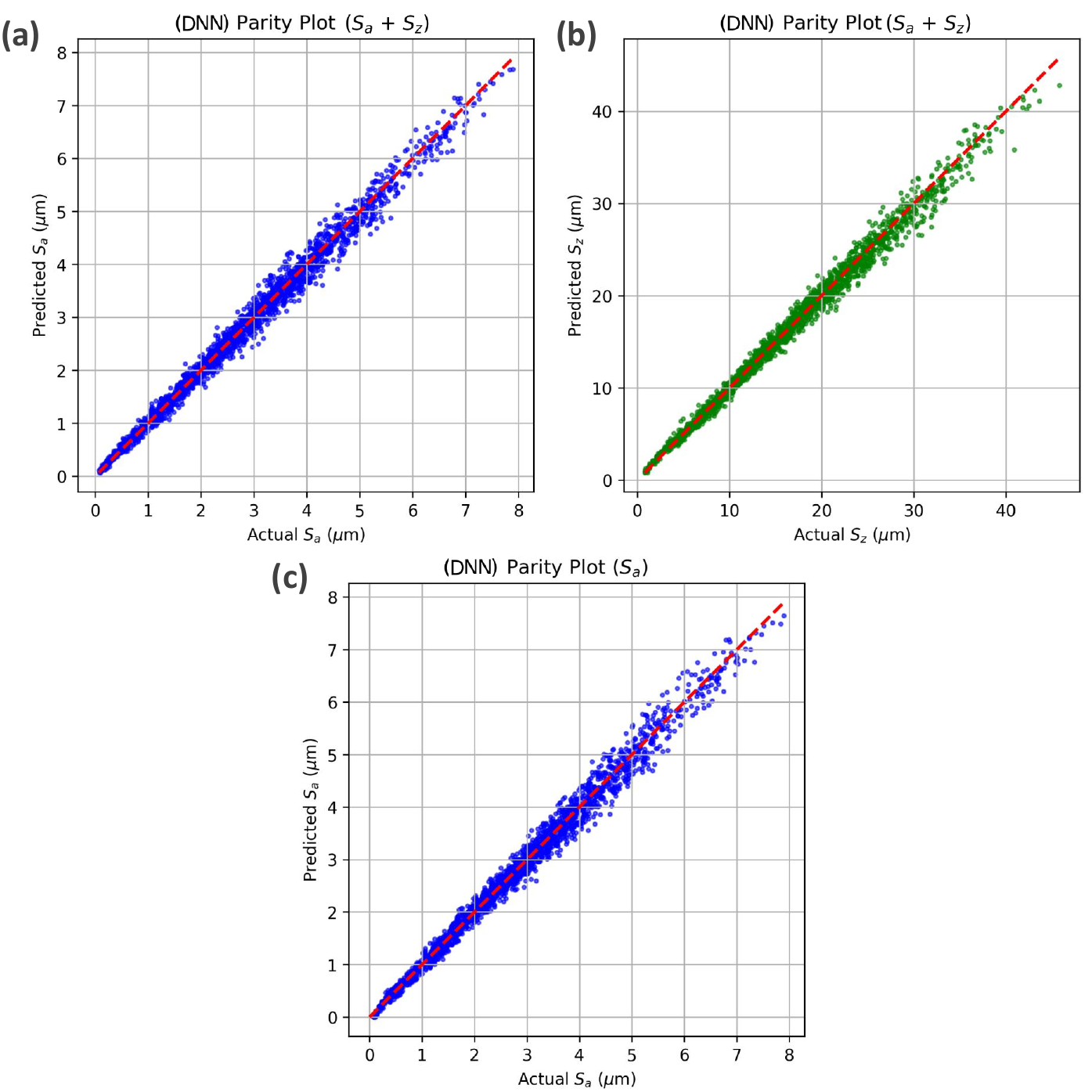}
                \caption{(DNN) Parity plots illustrating the agreement between predicted and actual values for the two-output scenario: (a) $S_a$ and (b) $S_z$, and for (c) the single-output scenario.}\label{Parity_Plots_DNN}
            \end{figure}

            The predictive performance is further confirmed by the parity plots shown in \autoref{Parity_Plots_DNN}. The parity plots illustrate the agreement between predicted and actual values from the dataset, where each point represents a single sample. Points closely clustered along the diagonal line indicate accurate predictions across the full range of target values. For both scenarios, the parity plots show that for lower values of $S_a$ and $S_z$, points are more tightly clustered along the diagonal compared to higher values, indicating that the model is more robust in predicting lower roughness values. This observation is supported by the distribution of the dataset, as shown in \autoref{Dataset_Multiplicity}a, where a larger number of samples are concentrated in the lower roughness range. Consequently, the model performs more accurately in regions with higher data density. Additionally, as seen in \autoref{Parity_Plots_DNN}c for the single-output scenario, points are more sparsely distributed, consistent with the slightly higher MSE observed during training.
     
            Together, these results demonstrate that the model reliably reproduces surface topography metrics while maintaining generalization on unseen data in the validation dataset.

    \subsection{Random forest (RF) model}
	    \label{subsec_random_forest_model}

        \subsubsection{RF architecture}
            \label{subsubsec_RF_architecture}

            In addition to the DNN, a Random Forest (RF) regression model was employed to capture the relationship between the input features and surface roughness metrics. RF is an ensemble-based, non-parametric learning algorithm that constructs a collection of decision trees and aggregates their predictions to improve generalization and reduce variance \cite{breiman2001random}. Each tree is trained on a randomly sampled subset of the data, and the final prediction is obtained by averaging the outputs of all trees. This approach reduces variance and enhances generalization compared to a single decision tree.

            To train the RF model, the same dataset used for the DNN was employed, with identical preprocessing steps, including data normalization and the partitioning of the dataset into training and validation sets. Although RF models are generally insensitive to feature scaling, normalization was applied to ensure a consistent preprocessing pipeline and to enable a fair comparison with the DNN results. Both single-output and two-output scenarios, consistent with the DNN procedure, were examined.

        \subsubsection{RF hyperparameter optimization}
            \label{subsubsec_RF_hyperparameter}

            The process of finding the optimal configuration for the RF model involves selecting the best combination of hyperparameters, including the number of trees ($n_{\text{estimators}}$), maximum depth, minimum samples split, minimum samples per leaf, maximum features, and bootstrap sampling, as listed in \autoref{tab_hyperparameters_RF}. These parameters collectively control the bias-variance trade-off, model complexity, and diversity of the ensemble.

           \begin{table}[!t]
                \centering
                \fontsize{8}{13}\selectfont
                \caption{Range of hyperparameters explored for determining the optimal RF architecture.}
                \label{tab_hyperparameters_RF}
                \begin{tabularx}{0.8\textwidth}{
                    >{\raggedright\arraybackslash}X
                    >{\raggedright\arraybackslash}X
                    >{\raggedright\arraybackslash}X
                    >{\raggedright\arraybackslash}X}
                
                    \toprule
                    Hyperparameter & Range / Values & Optimized value (two-output) & Optimized value (single-output) \\ \toprule
        
                    Number of trees ($n_{\text{estimators}}$)  & 100 -- 1000 & 799 & 433 \\
        
                    Maximum depth (\textit{max\_depth}) & 5 -- 50 & 33 & 40 \\
        
                    Minimum samples split (\textit{min\_samples\_split}) & 2 -- 20 & 3 & 4 \\
        
                    Minimum samples per leaf (\textit{min\_samples\_leaf}) & 1 -- 10 & 1 & 2 \\

                    Maximum features (\textit{max\_features}) & Sqrt, log2, None & None & None \\

                    Bootstrap sampling (\textit{bootstrap}) & True, false & True & True \\
         
                    \bottomrule
                \end{tabularx}
            \end{table}

            Both two-output and single-output scenarios were investigated. In the two-output case, unlike the single-output model, the algorithm constructs decision trees that minimize the joint variance across both outputs, effectively capturing the interdependencies between target variables. Similar to the DNN optimization process, each trial in the RF hyperparameter search involved training a candidate model on the training set and evaluating its performance on the validation set. The objective function was defined as the MSE averaged across both outputs. The optimization process was further enhanced using Optuna’s Median Pruner, which terminates underperforming trials early based on intermediate results. A total of 50 trials were conducted.

\begin{figure}[!t]%
                \centering
                \includegraphics[width=0.9\textwidth]{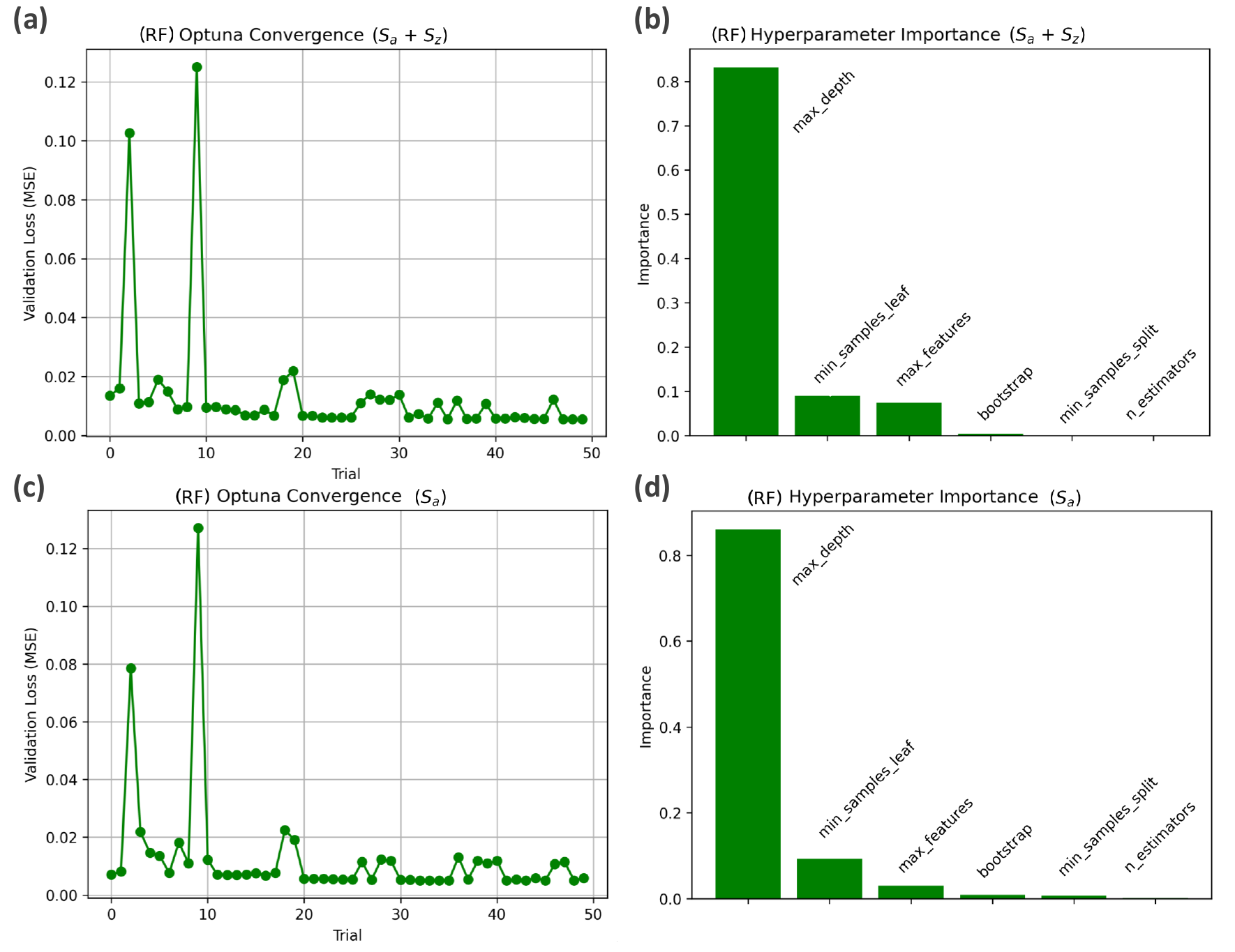}
                \caption{(RF) Convergence of the hyperparameter optimization process over successive trials for (a) two-output scenario and (c) one-output scenario. Relative contribution of each hyperparameter to the observed variation in validation loss for (b) two-output scenario and (d) one-output scenario.}\label{Hyperparameter_Importance_RF}
            \end{figure}

            The convergence behavior for the two-output and single-output scenarios is shown in \autoref{Hyperparameter_Importance_RF}a and \autoref{Hyperparameter_Importance_RF}c, respectively. Both cases converge to stable values, with the lowest MSE observed at 0.0055 for the two-output scenario and 0.005 for the single-output case. Among the hyperparameters, the maximum depth was the most influential for both scenarios. Parameters such as minimum samples per leaf and maximum features had a moderate effect, while the remaining hyperparameters exhibited minimal impact on model performance, as illustrated in \autoref{Hyperparameter_Importance_RF}b and \autoref{Hyperparameter_Importance_RF}d.

            The best-performing hyperparameters are listed in \autoref{tab_hyperparameters_RF}. A significant difference is observed in the number of trees between the two scenarios, with the two-output model requiring nearly twice as many trees as the single-output case. This can be attributed to the increased complexity of the learning task, as the model must simultaneously capture relationships for multiple target variables. Consequently, a larger ensemble is needed to adequately reduce variance and improve generalization performance. Interestingly, the single-output model exhibits a greater maximum depth compared to the two-output scenario. This suggests that, while fewer trees are sufficient, each tree must grow deeper to fully capture the intricate patterns of a single target.

        \subsubsection{RF model performance}
            \label{subsubsec_RF_performance}

            Following the hyperparameter optimization process, the corresponding configuration identified for each scenario was used to train the model. During training, model performance was evaluated using the MSE per output. For the two-output scenario, the training MSE is 0.00111 and the validation MSE is 0.00632, whereas for the single-output scenario, the training MSE is 0.00119 and the validation MSE is 0.00507. This indicates that the training losses are comparable between scenarios; however, the single-output model outperforms the two-output model on the validation set.

\begin{figure}[!t]%
                \centering
                \includegraphics[width=0.9\textwidth]{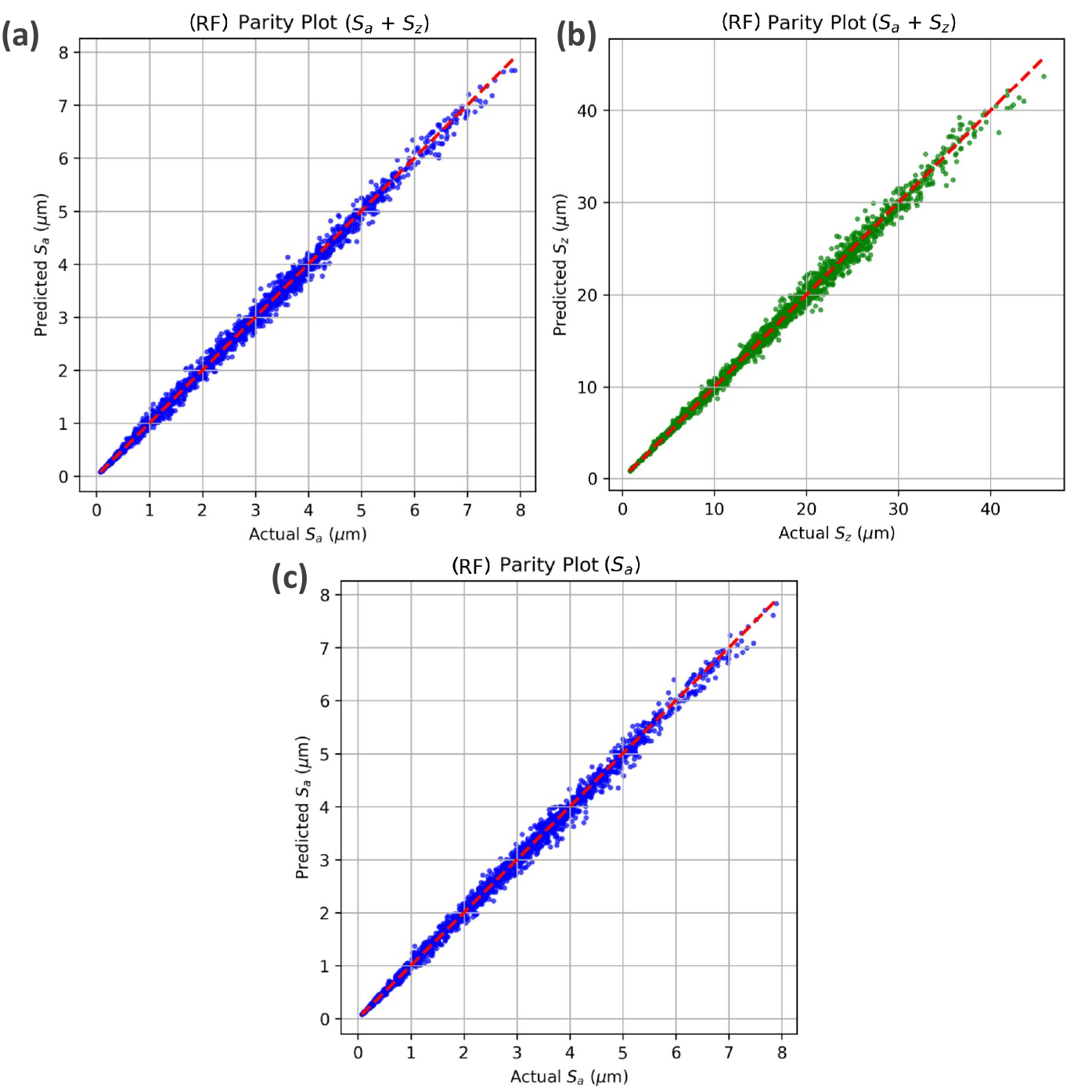}
                \caption{(RF) Parity plots for the two-output scenario: (a) $S_a$ and (b) $S_z$, and for (c) the single-output scenario.}\label{Parity_Plots_RF}
            \end{figure}

            Furthermore, \autoref{Parity_Plots_RF} shows the parity plots for both scenarios. The distribution of predicted points for $S_a$ is similar in both cases, exhibiting tight clustering at lower values and a slightly more scattered distribution around the diagonal at higher values. For $S_z$, the points are more dispersed, indicating higher prediction error and lower accuracy compared to $S_a$ (see \autoref{Parity_Plots_RF}b). When compared to the parity plots of the DNN models shown in \autoref{Parity_Plots_DNN}, the RF models display tighter point distributions with less sparsity, suggesting superior predictive performance. This observation is further supported by the lower MSE values obtained for the RF scenarios compared to the DNN training process given in \cref{subsubsec_DNN_performance}.

\section{Inverse design}
    \label{sec_inverse_design}

\begin{figure}[!t]%
        \centering
        \includegraphics[width=0.99\textwidth]{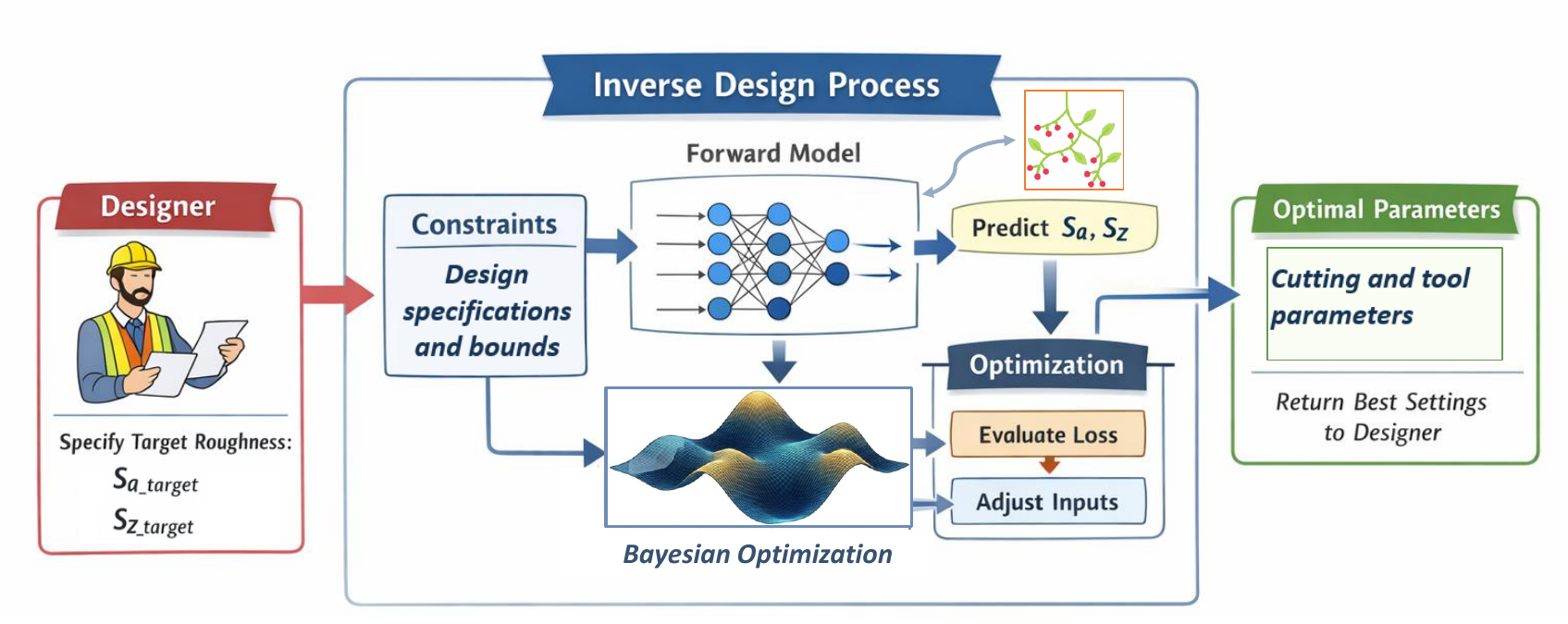}
        \caption{Optimization-based inverse design of the milling process.}\label{Inverse_Design}
    \end{figure}

    In the inverse design of the milling process considered in this study, the inherent nature of the dataset, which is characterized by a many-to-one relationship, necessitates the use of forward-trained ML models within an optimization pipeline, as shown in \autoref{Inverse_Design}. The core idea is that the designer specifies the desired average roughness and maximum height values as target requirements, and the framework subsequently provides multiple feasible combinations of input parameters that yield the desired surface roughness response. Additionally, these requirements can be further refined by incorporating additional design constraints such as tool selection or cutting conditions.

    As shown in \autoref{Inverse_Design}, the trained DNN or RF models serve as fast and differentiable proxies, enabling efficient exploration of the design space without the need for repeated experiments or high-fidelity simulations. The inverse design problem is thus formulated as an optimization task in which the objective is to minimize the deviation between the predicted and target roughness metrics. Depending on the model, both single-output and multi-output formulations can be incorporated. The implementation of the inverse deign framework was carried out using \texttt{optuna} and is available on GitHub at \url{https://github.com/HadiBakhshan/roughness-inverse-ml.git}.

    \subsection{Bayesian optimization (BO)}
	    \label{subsec_bayesian_optimization}

        The optimization procedure was carried out using BO, which serves as an efficient method for solving inverse design problems in high-dimensional and non-convex search spaces \cite{snoek2012practical}. BO operates by constructing a probabilistic surrogate model of the objective function and then selecting new sampling points based on an acquisition strategy that balances exploration and exploitation. In this work, the Tree-structured Parzen Estimator (TPE) \cite{watanabe2023tree} was employed. This approach samples new candidate solutions from regions with a high probability of improvement by modeling the distributions of good and bad observations separately.

        In BO, the optimization process begins with an initial set of randomly sampled trials to explore the global design space. As more evaluations are performed using the machine learning model, the TPE sampler progressively refines its probabilistic model and increasingly focuses on regions that yield lower loss values. Consequently, the optimizer converges toward promising regions of the search space. However, BO has a known limitation in its tendency toward exploitation, where it repeatedly samples near previously identified optima. This behavior can lead to a localized search. In the present case, where the goal is to identify a broad set of optimal design solutions and combinations of cutting and tool parameters, this results in solutions clustered around a single configuration.

        To address this issue, three measures were implemented. First, the discrete parameters, namely the number of inserts and their radii, were removed from the continuous design space and explicitly enumerated. Separate optimization runs were then conducted for each combination. This ensures that the optimizer explores all physically meaningful tool configurations rather than converging prematurely to a single dominant configuration, while focusing the optimization on the continuous input parameters of the forward model. Second, to discourage repeated sampling of nearby solutions, a diversity penalty was incorporated into the objective function. This encourages broader exploration of the continuous parameter space and helps identify multiple distinct solutions within each configuration. Third, the number of startup trials for the TPE sampler was increased, leading to more extensive random exploration before the exploitation phase begins.

    \subsection{Inverse design results and analysis}
	    \label{subsec_inverse_design_result}
        
        In this section, we define two target case studies to evaluate the inverse design capability developed. The objective is to provide the $S_a$ and $S_z$ values to the framework and obtain possible solutions from the design space using both DNN and RF prediction models. Only two-output prediction models are considered here, as the procedure is identical for single-output scenarios, such as case studies where the average surface roughness ($S_a$) is the sole target.

        \subsubsection{Case 1: Lower roughness regime}
            \label{subsubsec_case_one_lower}

            In the first example analysis, low values for both $S_a$ and $S_z$ are selected, as the data density is higher in the lower range compared to the higher values of the roughness metrics, as shown in \autoref{Dataset_Multiplicity}a. It should be noted that, considering the output space distributions illustrated in \autoref{Dataset_Multiplicity}a, the selected target values must be feasible and lie within the dataset distribution. Therefore, for the low-value case study, the target values are set to 2 and 10 for $S_a$ and $S_z$, respectively.

\begin{figure}[!t]%
                \centering
                \includegraphics[width=0.99\textwidth]{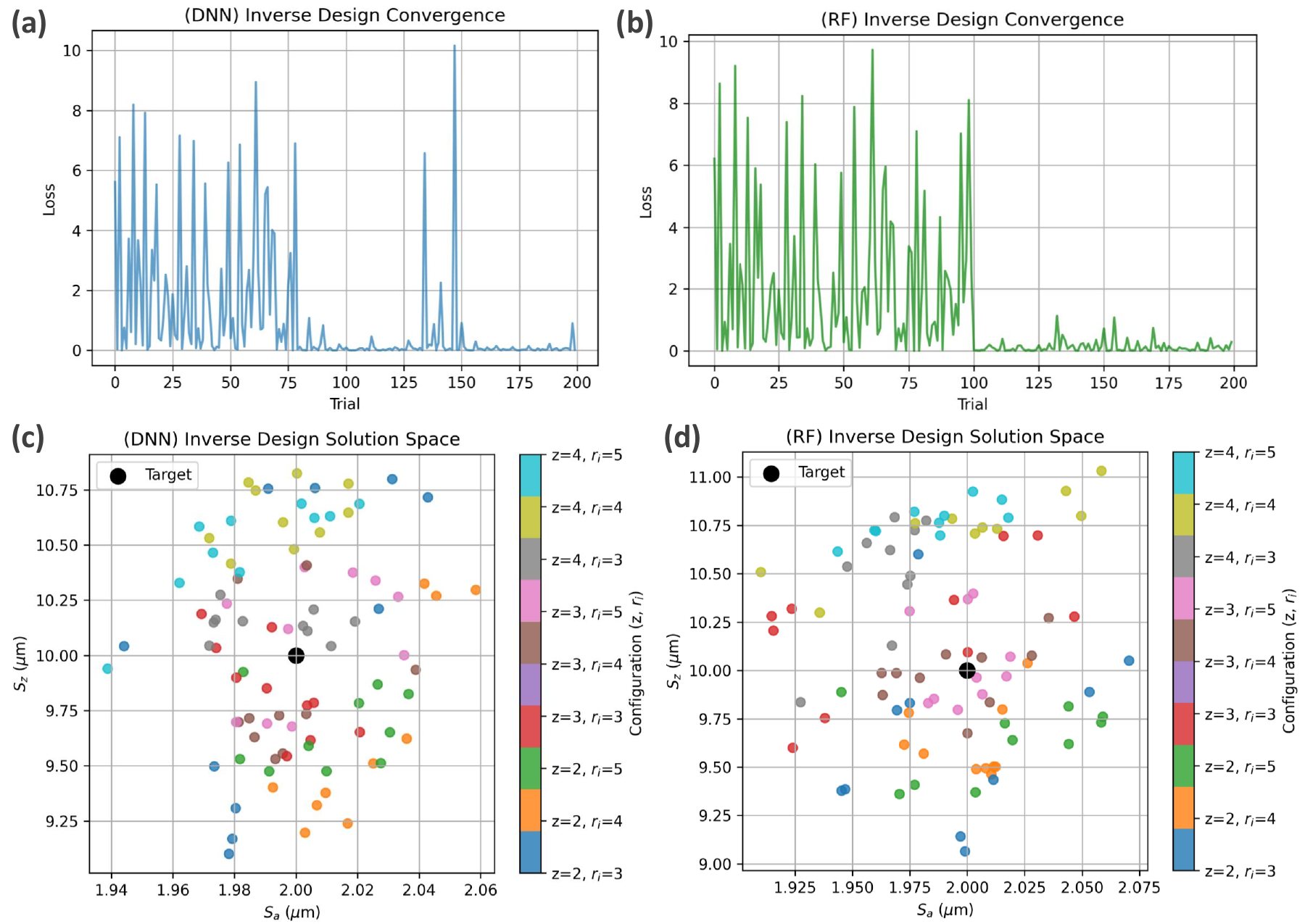}
                \caption{Case 1: Inverse design convergence for (a) DNN and (b) RF models, along with their corresponding design solution spaces for (c) DNN and (d) RF models.}\label{Case_1_Inverse_Solution}
            \end{figure}

            \autoref{Case_1_Inverse_Solution} presents the convergence plots and solution spaces for both the DNN and RF models. A total of 200 trials were selected for the optimization search process. As shown in \autoref{Case_1_Inverse_Solution}a, the corresponding loss values for the DNN indicate convergence after approximately 75 trials; however, higher loss values reappear around trial 150.
            
            This behavior differs from the RF convergence trend, where, as illustrated in \autoref{Case_1_Inverse_Solution}b, the model converges after approximately 100 trials and subsequently becomes stable. Such delayed convergence is not typically observed in standard BO, which is generally efficient at converging to local optima. However, the methodology adopted here, looping over discrete values of the number of inserts and their radii, is necessary to perform a more comprehensive exploration of the solution space. This approach helps avoid entrapment in local minima and prevents the search from making only minimal forward and backward adjustments in continuous variables, where changes may be insignificant.
            
            The target values of $S_a = 2$ and $S_z = 10$ are located near the center of the solution space for both DNN and RF models, as shown in \autoref{Case_1_Inverse_Solution}c and d, respectively. The trials are distributed around these target points, with feasible solutions corresponding to lower loss values highlighted in different colors. These solutions are clustered closer to the target, indicating that the search process begins from distant regions with higher loss values and progressively converges toward regions with lower loss.
            
            It can also be observed that, for both DNN and RF solution spaces, the recommended solutions consist of various combinations of the number of inserts and their corresponding radii, reflecting multiple feasible design alternatives.

\begin{figure}[!t]%
                \centering
                \includegraphics[width=0.85\textwidth]{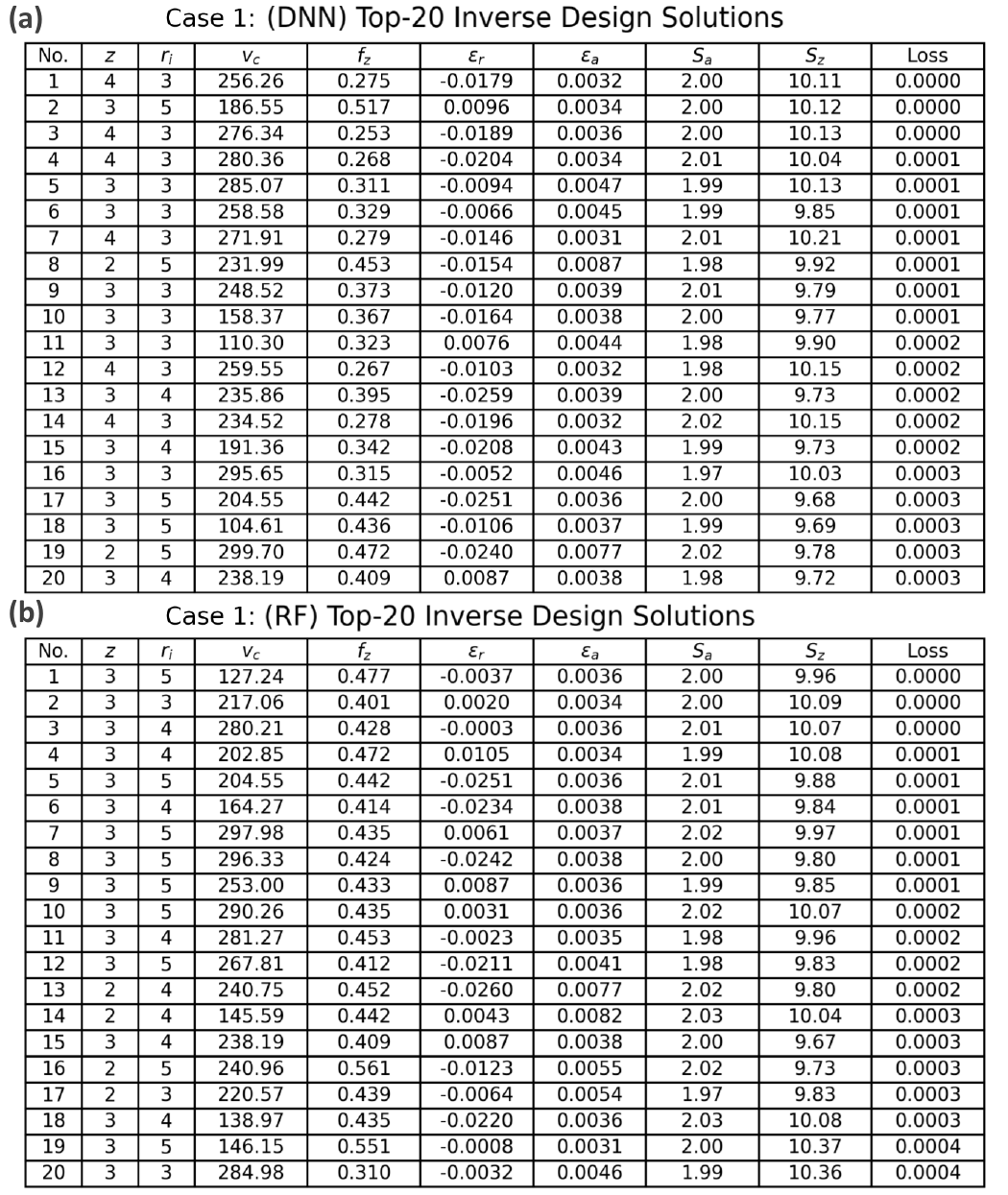}
                \caption{Case 1: Top 20 candidates with the best performance, based on the lowest loss values obtained in the inverse design process, for (a) DNN and (b) RF models.}\label{Case_1_Table}
            \end{figure}

            To better evaluate the performance of the inverse design pipelines for both models in Case 1, the top 20 candidates, ranked based on the lowest obtained loss values, are presented in \autoref{Case_1_Table}. The table lists the predicted cutting and tool parameters corresponding to the given target surface roughness metrics. It can be observed that, for various combinations of the number of inserts and insert radii, the proposed top candidates yield predicted $S_a$ and $S_z$ values that are closely aligned with the target values.
            
            It is worth noting that the predicted ranges of the cutting and tool parameters are constrained within the design input space used for training the forward DNN and RF models. The listed candidates also provide insight into the design flexibility in the low roughness regime, where a diverse set of feasible solutions with varying cutting and tool parameters is available.

\begin{figure}[!t]%
                \centering
                \includegraphics[width=0.99\textwidth]{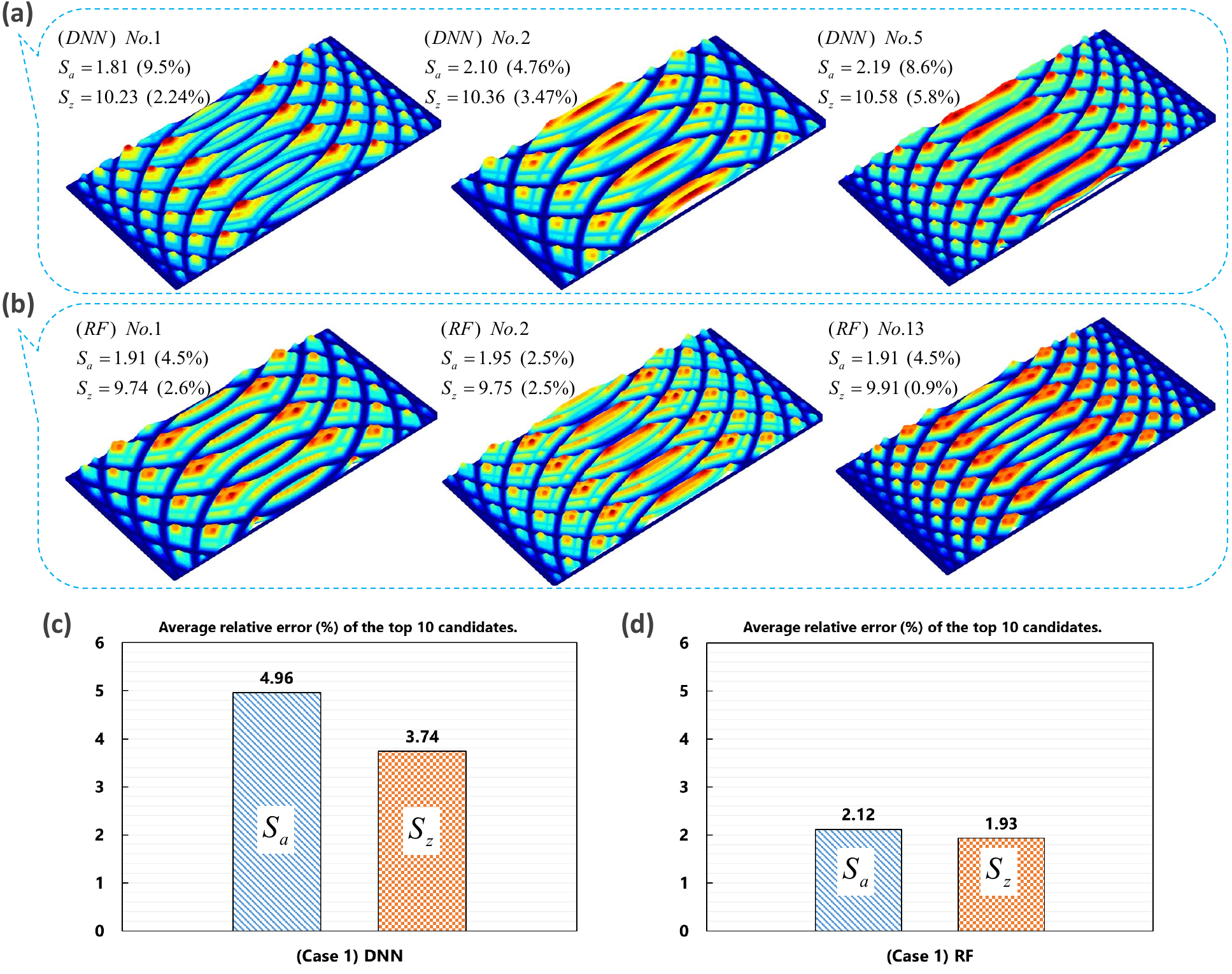}
                \caption{Case 1: Selected candidates for surface topography demonstration, along with the relative prediction errors of $S_a$ and $S_z$ for (a) DNN and (b) RF models. The average relative error of the top 10 candidates obtained from the inverse design process for $S_a$ and $S_z$ is shown for (c) DNN and (d) RF models.}\label{Case_1_Topographies}
            \end{figure}

            \autoref{Case_1_Topographies} illustrates selected candidates from the top 20 design solutions, chosen to represent different combinations of the number of inserts and insert radii. These selections enable visualization of the variation in surface topography patterns resulting from changes in the recommended input parameters. The topographies are generated by feeding the predicted cutting and tool parameters from \autoref{Case_1_Table} into the computational framework used for dataset generation. This framework provides the reference values of $S_a$ and $S_z$, enabling evaluation of the proposed inverse design solutions.
            
            The relative errors between the inverse design candidates and the computational reference results are shown in \autoref{Case_1_Topographies}a and b for the DNN and RF models, respectively. As illustrated in \autoref{Case_1_Topographies}a, candidates 1, 2, and 5 from the DNN model are selected, and the errors in $S_a$ and $S_z$ indicate that the predictions for $S_z$ generally exhibit lower error compared to those for $S_a$.

            \autoref{Case_1_Topographies}b presents three candidates from the RF model predictions, where lower errors are observed for both $S_a$ and $S_z$ compared to those obtained from the DNN model. Overall, the RF model demonstrates superior predictive accuracy for the selected candidates. 

            To further support this observation, the average errors for the top 10 candidates listed in \autoref{Case_1_Table} for both DNN and RF models are calculated and presented in \autoref{Case_1_Topographies}c and d, respectively. The predicted cutting and tool parameter configurations are evaluated using the same computational framework employed for dataset generation to compute the corresponding roughness metrics, which are then compared against the target values.
            
            The average relative error of $S_a$ is 4.96\% for the DNN model and 2.12\% for the RF model, indicating that the RF model provides superior predictive performance. A similar trend is observed for $S_z$, where the RF model achieves an average error of 1.93\%, compared to 3.74\% for the DNN model. These error ranges are encouraging for the computational benchmark considered and suggest that the inverse design pipeline is consistent within the studied synthetic design space. Furthermore, the RF model shows improved performance in Case 1, which corresponds to the lower roughness region of the dataset space.

        \subsubsection{Case 2: Higher roughness regime}
            \label{subsubsec_case_two_higher}

            The target values for $S_a$ and $S_z$ are selected from the higher range of the dataset distribution, where the data is more scattered and sparse, as shown in \autoref{Dataset_Multiplicity}a. In these regions, a lower versatility of design combinations is expected, since the data is limited and unevenly distributed, and only specific combinations contribute to generating such high roughness metrics. For Case 2, the target values are set to $S_a = 6$ and $S_z = 33$.
            
\begin{figure}[!t]%
                \centering
                \includegraphics[width=0.99\textwidth]{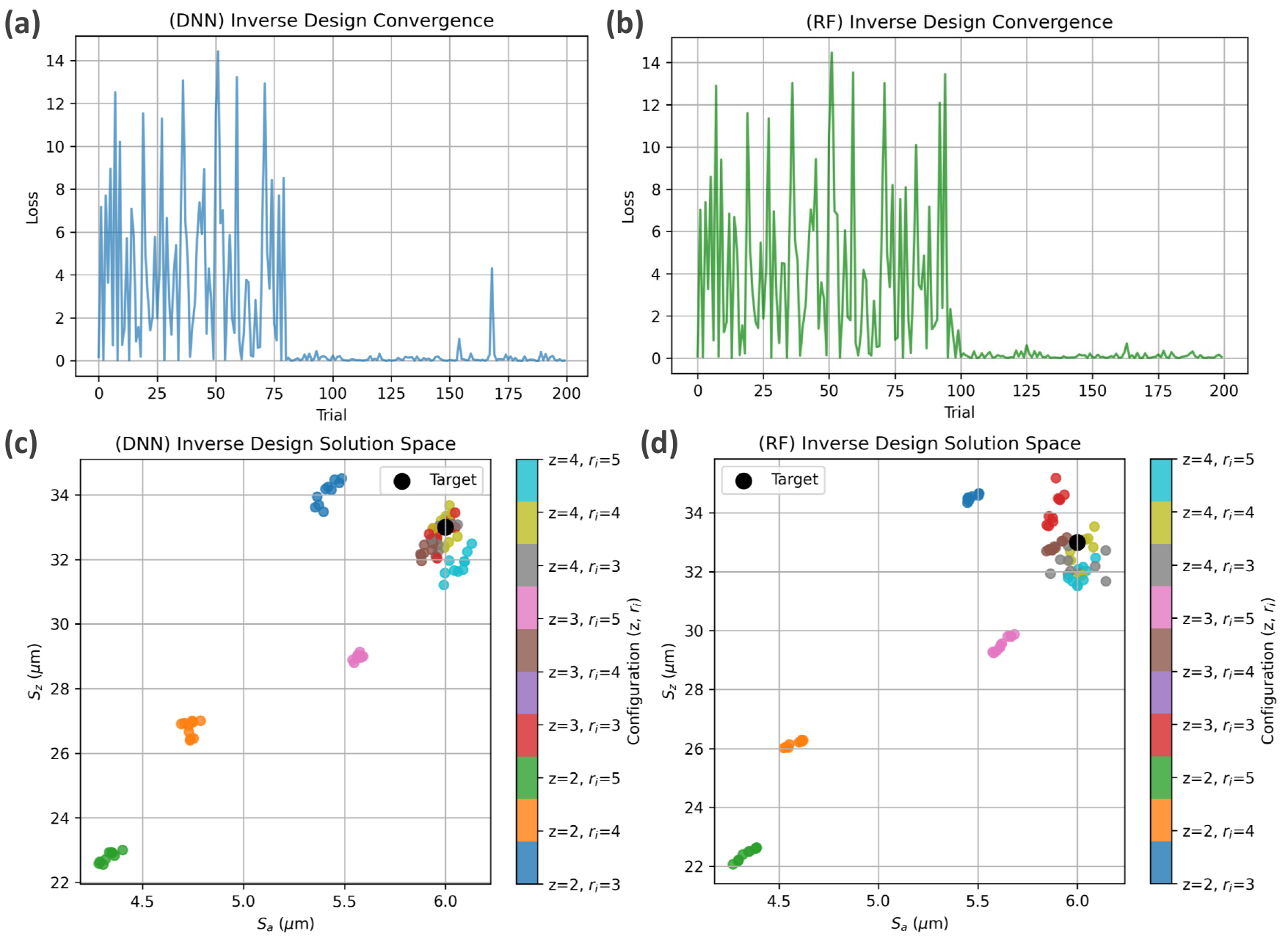}
                \caption{Case 2: Inverse design convergence for (a) DNN and (b) RF models, along with their corresponding design solution spaces for (c) DNN and (d) RF models.}\label{Case_2_Inverse_Solution}
            \end{figure}

            The convergence plots of the inverse design process for the DNN and RF models are presented in \autoref{Case_2_Inverse_Solution}a and b, respectively. Similar convergence trends can be observed in both models. For the DNN model, convergence occurs after approximately 75 trials, although some fluctuations in the loss values are still present. In contrast, the RF model demonstrates more robust behavior and, similar to Case 1, achieves stable convergence after approximately 100 trials. The higher loss values observed in the early trials are associated with design points that are widely scattered across the solution space, where larger deviations from the target values naturally result in higher loss.

            Furthermore, the distribution pattern of the design points in Case 2 differs from that of Case 1, as illustrated in \autoref{Case_2_Inverse_Solution}c and d. Overall, the design points appear clustered in certain regions that are either far from or close to the target values. As shown in \autoref{Case_2_Inverse_Solution}c, which represents the solution space for the DNN model, different combinations of the number of inserts and their radii are indicated using distinct colors. Some design combinations are not feasible for achieving the target roughness metrics and are located far from the target region. For example, the combinations $z = 2, r_i = 5$ (green) and $z = 2, r_i = 4$ (orange) are positioned in regions corresponding to lower values of $S_a$ and $S_z$, making them unsuitable for this case. These clusters indicate that certain combinations of insert number and radius are inherently limited in their ability to reach higher roughness values.

            \begin{figure}[!t]%
                \centering
                \includegraphics[width=0.85\textwidth]{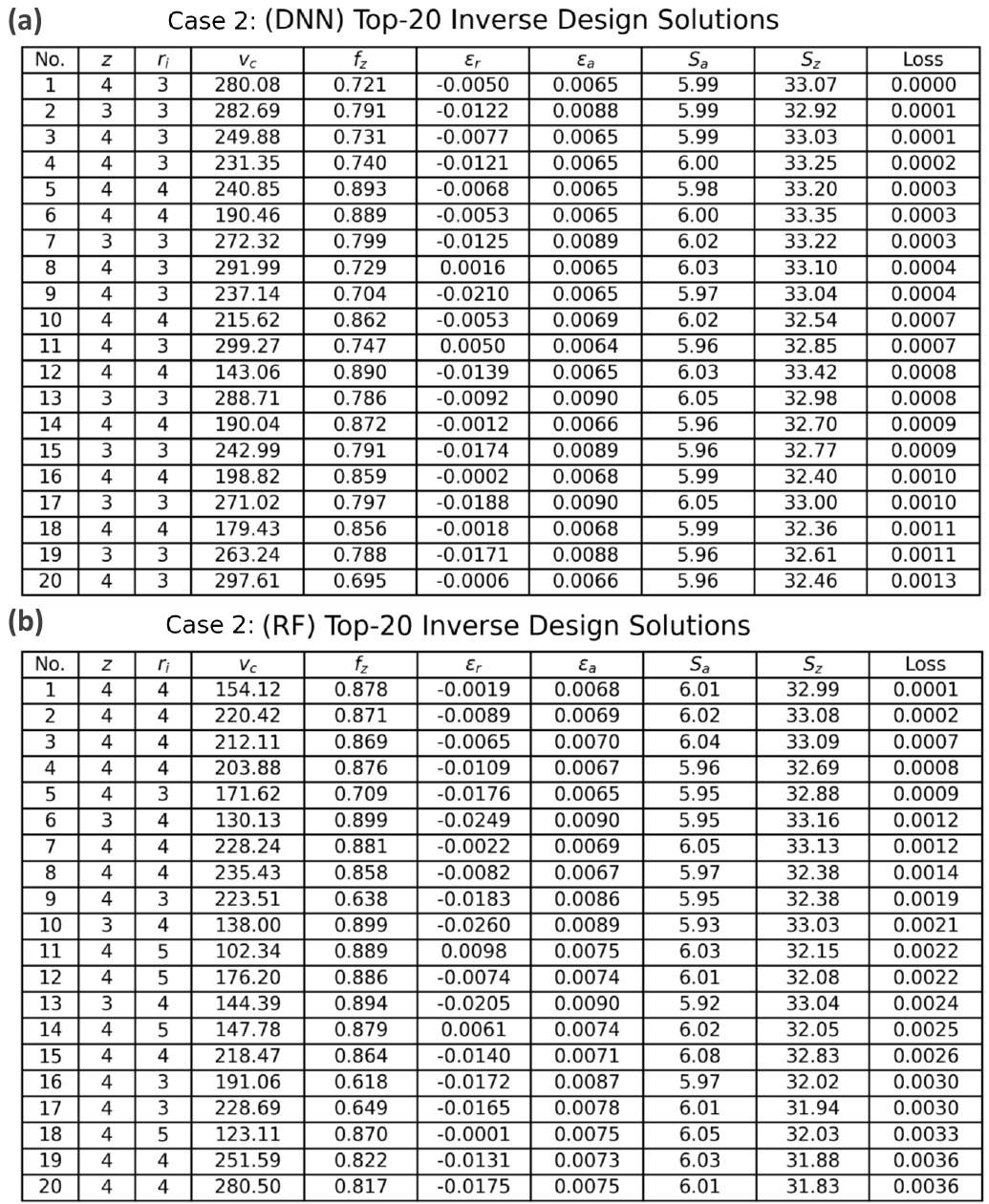}
                \caption{Case 2: Top 20 candidates with the best performance, based on the lowest loss values obtained in the inverse design process, for (a) DNN and (b) RF models.}\label{Case_2_Table}
            \end{figure}

            \begin{figure}[!t]%
                \centering
                \includegraphics[width=0.99\textwidth]{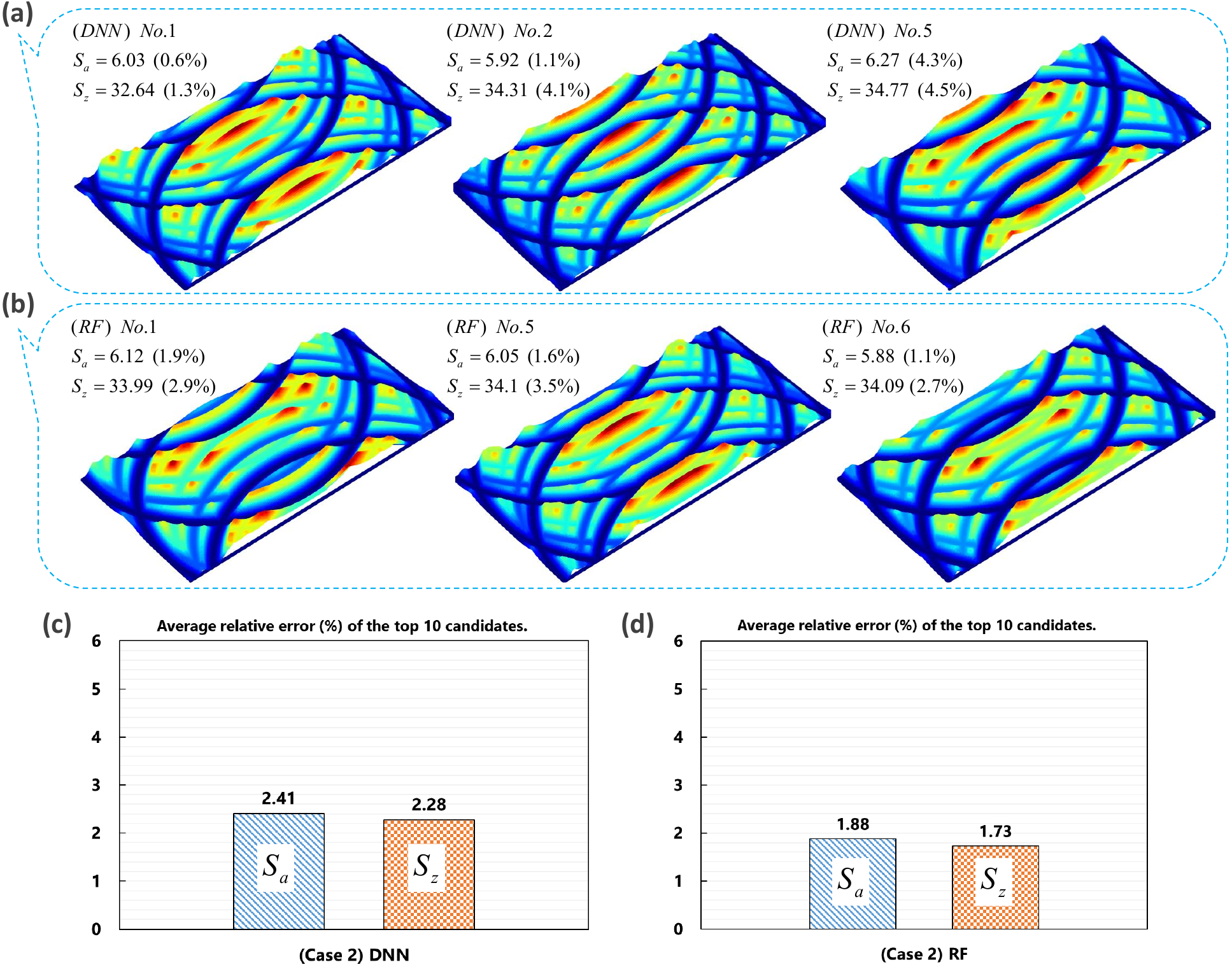}
                \caption{Case 2: Selected candidates for surface topography demonstration, along with the relative prediction errors of $S_a$ and $S_z$ for (a) DNN and (b) RF models. The average relative error of the top 10 candidates obtained from the inverse design process for $S_a$ and $S_z$ is shown for (c) DNN and (d) RF models.}\label{Case_2_Topographies}
            \end{figure}

            A similar behavior is observed in the RF solution space, indicating consistent design space characteristics across both models. Additionally, a limited number of feasible combinations are locally distributed around the target region, resulting in a more localized convergence process. This behavior is likely due to the models being trained on relatively sparse data in this region, which reduces the diversity of viable design combinations in terms of the number of inserts and their radii. 

            Similar to Case 1, the top 20 candidates with the best performance, based on the lowest loss values, are presented in \autoref{Case_2_Table}. From this set, selected candidates are used to illustrate the corresponding surface topographies, as shown in \autoref{Case_2_Topographies}a and b for the DNN and RF models in Case 2, respectively.
            
            The relative errors are also low and within an acceptable range. For the top 10 candidates identified for each of the DNN and RF models, the average relative errors are presented in \autoref{Case_2_Topographies}c and d, respectively. These values are lower compared to those observed in Case 1 for both $S_a$ and $S_z$.
            
            The average relative error between the inverse design solutions and the computational reference results, across both low and high roughness regimes of the dataset, indicates that the models are capable of predicting and proposing acceptable cutting and tool parameters in both densely populated and sparse regions of the design space. This demonstrates the applicability and robustness of the inverse design framework, with error levels remaining within an acceptable range of below 5\%.

            It is worth noting that, in this study, two cases were investigated without imposing additional design constraints or requirements. However, the optimization framework is flexible and can accommodate conditional or soft constraints. By specifying particular values or ranges for tool or machining parameters, the optimization search space can be guided toward desired regions. For example, a designer may choose to neglect run-out errors. By incorporating this assumption as a constraint in the optimization process, a penalty can be applied to the loss function whenever the condition is violated. This approach enhances the flexibility and practical applicability of the framework, providing designers with greater control over the solution space.


\section{Conclusions}
    \label{conclusions}

    In this paper, a surface roughness-based framework is developed for the inverse design of the milling process using ML and BO. The procedure operates by providing target surface roughness metrics, such as average roughness or maximum surface height. The trained ML models, including a DNN and a RF, are then employed within a BO algorithm to generate feasible solutions from the design space of milling process and tool parameters. The main conclusions can be summarized as follows:

    \begin{enumerate}

    \item To augment the dataset, the influential parameters in the computational framework include cutting speed and feed rate from the process parameters, as well as the number of inserts, insert radius, and run-out errors from the tool parameters. The outputs considered are average roughness and maximum height roughness.
    
    \item The dataset exhibits multiplicity behavior, indicating a many-to-one mapping. This characteristic prevents direct training of ML models that use surface roughness as input and process and tool parameters as outputs.
    
    \item Both DNN and RF models perform adequately, achieving low MSE values in validation dataset analyses for both single-output and multi-output scenarios.
    
    \item The inverse design is conducted by incorporating BO to effectively utilize the trained ML models, thereby narrowing the design space and proposing optimal design configurations from all possible candidates.
    
    \item The overall average relative errors between the top-scored design candidates and their corresponding actual data are low, typically below 5\%, indicating the robustness of the proposed inverse design methodology.
    
    \end{enumerate}

    The inverse design framework has demonstrated robust capability in predicting machining and tool parameters based on trained ML models. These predictive models can be further improved by incorporating additional influential parameters affecting surface roughness performance, including workpiece properties, cutting forces, and vibration characteristics. However, this advancement requires more efficient computational approaches for dataset generation, or alternatively, the use of few-shot learning techniques based on experimental results within physics-informed frameworks.
    
    Moreover, the entire inverse design workflow including dataset augmentation, model development and training, the inverse design pipeline, and design specification recommendations can be further enhanced and orchestrated using agent-based systems within the emerging paradigm of agentic AI. Such an approach could enable interactive, real-time user engagement through conversational interfaces, thereby facilitating and streamlining the design process.


\section*{Declaration of competing interest}
	
   The authors declare that they have no known competing financial interests or personal relationships that could have appeared to influence the work reported in this paper.




\section*{Data availability}
    The data for this paper are available on GitHub \url{https://github.com/HadiBakhshan/roughness-inverse-ml.git}.

\section*{Code availability}

The code for this paper is available on GitHub \url{https://github.com/HadiBakhshan/roughness-inverse-ml.git}.


	\bibliographystyle{elsarticle-num} 
	\bibliography{Bibliography}

\end{document}